\documentclass[letterpaper, 12pt]{article}
\usepackage{graphicx} 
\usepackage{natbib}
\usepackage{caption}
\usepackage{multirow}
\usepackage{amssymb}
\title{The Smallest Particles in Saturn's A and C Rings}
\author{Rebecca A. Harbison\footnote{Department of Astronomy, Cornell University, Ithaca, NY, 14853} \footnote{Corresponding author} , Philip D. Nicholson$^{\star}$ \& Matthew M. Hedman$^{\star}$} 
\begin{document}
\maketitle
\begin{abstract}
Radio occultations of Saturn's main rings by spacecraft suggest a power law particle size-distribution down to sizes of the order of 1 cm \citep{1983Marouf1}, \citep{1985Zebker1}. The lack of optical depth variations between ultraviolet and near-IR wavelengths indicate a lack of micron-sized particles. Between these two regimes, the particle-size distribution is largely unknown. A cutoff where the particle-size distribution turns over must exist, but the position and shape of it is not clear from existing studies.

Using a series of solar occultations performed by the VIMS instrument on-board Cassini in the near-infrared, we are able to measure light forward scattered by particles in the A and C rings. With a model of diffraction by ring particles, and the previous radio work as a constraint on the slope of the particle size distribution, we estimate the minimum particle size using a truncated power-law size distribution. The C Ring shows a minimum particle size of $4.1^{+3.8}_{-1.3}$ mm, with an assumed power law index of $q=3.1$ and a maximum particle size of 10 m. 

The A Ring signal shows a similar level of scattered flux, but modeling is complicated by the presence of self-gravity wakes, which violate the assumption of a homogeneous ring, and higher optical depths, which require multiple-order scattering.   If $q<3$, our A Ring model requires a minimum particle size below one millimeter ($< 0.34$ mm for an assumed $q=2.75$, or $0.56^{+0.35}_{-0.16}$ mm for a steeper $q=2.9$) to be consistent with VIMS observations.  These results might seem to contradict previous optical \citep{1993Dones1} and infrared \citep{2000French1} work, which implied that there were few particles in the A Ring smaller than 1 cm.  But, because of the shallow power law, relatively little optical depth (between 0.03 and 0.16 in extinction, or 0.015 - 0.08 in absorption) is provided by these particles.  \end{abstract}

NOTICE: this is the authorÕs version of a work that was accepted for publication in \textit{Icarus}. Changes resulting from the publishing process, such as peer review, editing, corrections, structural formatting, and other quality control mechanisms may not be reflected in this document. Changes may have been made to this work since it was submitted for publication. A definitive version was subsequently published in \textit{Icarus}, [volume 226, issue 2, 11/2013] 10.1016/j.icarus.2013.08.015

\section{Introduction}
The vast majority of particles that make up Saturn's main rings cannot be seen individually, but as an aggregate they become one of the most striking objects in the Solar System.  From past observations, we know that the ring particles come in a variety of sizes. 

The Voyager radio science experiment used radio occultations at 3.6 and 13 cm to probe the ring particles by two methods.  Direct inversion of the radio signal forward-scattered by meter-sized particles produced a size distribution showing a sharp drop-off above a radius of $\sim$5 m, while the differential optical depth measured between the two bands used in the occultation allowed a power law to be fit between particle radii of 1 m and 1 cm \citep{1983Marouf1,1985Zebker1}.  However, the Voyager radio science experiment was insensitive to particles smaller than 1cm; smaller ring particles does not absorb even the shorter 3.6 cm radio waves from Voyager. 

\citet{2000French1} used the 28 Sagitarii (28 Sgr) stellar occultation, as observed from Earth in July, 1989 at wavelengths between 1 and 4 $\mu$m, to look for light forward-scattered by ring particles.  Unlike the monochromatic radio-science experiments, they could not separate forward-scattered light from light directly transmitted through the rings.  However, the differing range and acceptance angle between this and the Voyager PPS stellar occultation allowed a gross measurement of forward-scattering.  This measurement could then be modeled with a truncated power law.  The 28 Sgr occultation, like Voyager, had limited sensitivity to particles under 1 centimeter, but for a different reason: the scattering angles of such small material was larger than the photometric aperture size, so could not be measured. 

Previous radio occultation and stellar occultation experiments were thus most sensitive to particles in the centimeter to meter range.   This situation changed with the arrival of the Cassini spacecraft at Saturn in 2004.  As Saturn was near its northern winter solstice in 2004, the rings were more open than when Voyager observed them, reducing the effective optical depth as seen from Earth and increasing the signal to noise for occultations by dense rings.  In addition to the 3.6 and 13 cm radio bands, Cassini can also transmit at 1.3 cm.  Not only does a shorter wavelength probe smaller particle sizes, but three measurements of the optical depth at different wavelengths allow for more exact models to constrain both the effective minimum particle size and effective power law index.  The C ring minimum particle size was estimated at 4 mm, while the data for the A ring suggest a larger minimum particle size \citep{2008Marouf1}.  A full discussion of these results can be found in \citet{2009Cuzzi1}.

While some micrometer-sized particles have been seen within the main rings, they are either found in transient spoke features \citep{2010DAversa1, 2013Mitchell1}, probably dislodged from the surfaces of larger ring particles \citep{2006Mitchell1}, or are confined to narrow, diffuse regions of the rings such as the Encke Gap ringlets \citep{2007Hedman1} and the `Charming Ringlet' in the Laplace Gap \citep{2010Hedman1}.  Differential optical depth, phase-function, and microwave emissivity measurements all show that very little dust persists within the main rings on a large scale in both space and time \citep{1993Dones1, 2000French1, 2005Spilker1}.  Theoretical work by \citet{2012Bodrova1} also has shown that, under unperturbed main ring conditions, particles smaller than a few centimeters will adhere onto larger ring particles.  

When Cassini entered Saturn orbit in 2004, its wide range of orbital geometries not only allowed for multiple radio and (space-based) stellar occultations, but also permitted the first solar occultations by the rings to be observed. The Visual and Infrared Mapping Spectrometer (VIMS) onboard Cassini can accept light through a special solar port, which has the attenuation needed to safely observe the Sun with the VIMS detector array.  Given the 0.5 milliradian pixel size of VIMS and its near infrared (0.9 to 5.2 microns) spectral range, the VIMS data are most sensitive to the previously-unsampled size regime of 100 microns to a few millimeters.

 In this work, we will use the VIMS solar occultations to examine this neglected regime, with the goal of setting an effective minimum radius on the ring particle size distribution in different regions.  Following a description of the VIMS solar port and the data taken during solar occultations, we will present our method for reducing the solar port data and separating the component of light scattered at small angles from the direct solar image.  Once this component is separated, it can be compared to a model of particle diffraction to estimate an effective minimum particle size for the C ring.  This model is then refined to account for the self-gravity wakes and higher optical depths observed within the A ring -- which violate several simplifying assumptions made at first -- and applied to that ring.
  
\section{Data}
\subsection{Basic Processing}
\label{sec:processingintro}

As of February 2010, Cassini has observed eleven solar occultations of the rings: see Tables \ref{table:aringobserv} and \ref{table:cringobserv} for a list.  The procedure for observing solar occultations involves taking a series of 12 pixel by 12 pixel (6 x 6 milliradians) multispectral images of the area of the sky around the Sun using the VIMS solar port,  which has an attenuation on the order of $10^{5}$. The instrument's visible channel is turned off, as the visible-light spectra, even through the solar port, saturate within a few milliradians of the Sun.  Thus, data obtained through the solar port have a wavelength coverage of 0.9 to 5.2 $\mu$m.  A single VIMS `cube' of two spatial and one spectral dimensions is constructed pixel by pixel, using a 2D scanning mirror.  Each pixel has an exposure time of 40 ms, and approximately 5 cubes of 144 pixels each are obtained for every minute of the occultation.  Each occultation data set is thus a time series of cubes -- one temporal dimension, two spatial, and one spectral.  For full details of the VIMS instrument, see \citet{2004Brown1}.  

The onboard VIMS signal processing electronics perform automatic background subtraction.  At the end of each line of 12 pixels, VIMS takes a spectrum of the thermal background signal by closing off the spectrometer from outside light and taking a measurement. Four measurements of this dark spectrum are averaged together, then subtracted from the last four lines of pixels taken.  As a result, each cube has a slightly uneven background subtraction, as there is some shot-noise variance even after averaging over four measurements.  In most cases, these three background spectra are within a data number (DN) or two of one another\footnotemark, but a cosmic ray can hit the detector during a background measurement, producing an artifictially high background in one or more spectral channels.

\footnotetext{Raw VIMS spectra represent photo-electron counts, but are returned in scaled integer form as Data Numbers.  The instrumental gain was set such that the detector read noise is $\approx$ 1 DN, or $\sim$ 300 electrons.}

In order to correct this, the background was re-added to the signal, returning the data to its raw form, and then the median of the three dark current spectra recorded for each cube was used as the background instead.  The slight temperature change when Cassini moves into the rings' shadow lowers the dark current by approximately 2 DN.  Hence the dark background subtracted is slightly dependent on the position of Cassini, so further time-averaging of the background was not done.  

The cubes showing the unocculted Sun were used as a reference to define transmission of the rings, and all measurements are reported either in units of transmission or in 'raw' data numbers (DN), rather than absolute flux. The position of the Sun within the image varied by well under a single pixel in each occultation, making any variable response due to a slightly different beam path within the solar port or the spectrometer minimal.  

\subsection{Instrumental Effects}
\label{sec:processinginstrument}

\begin{figure} [htbp]
 \centering 
 \includegraphics[width=0.45\textwidth]{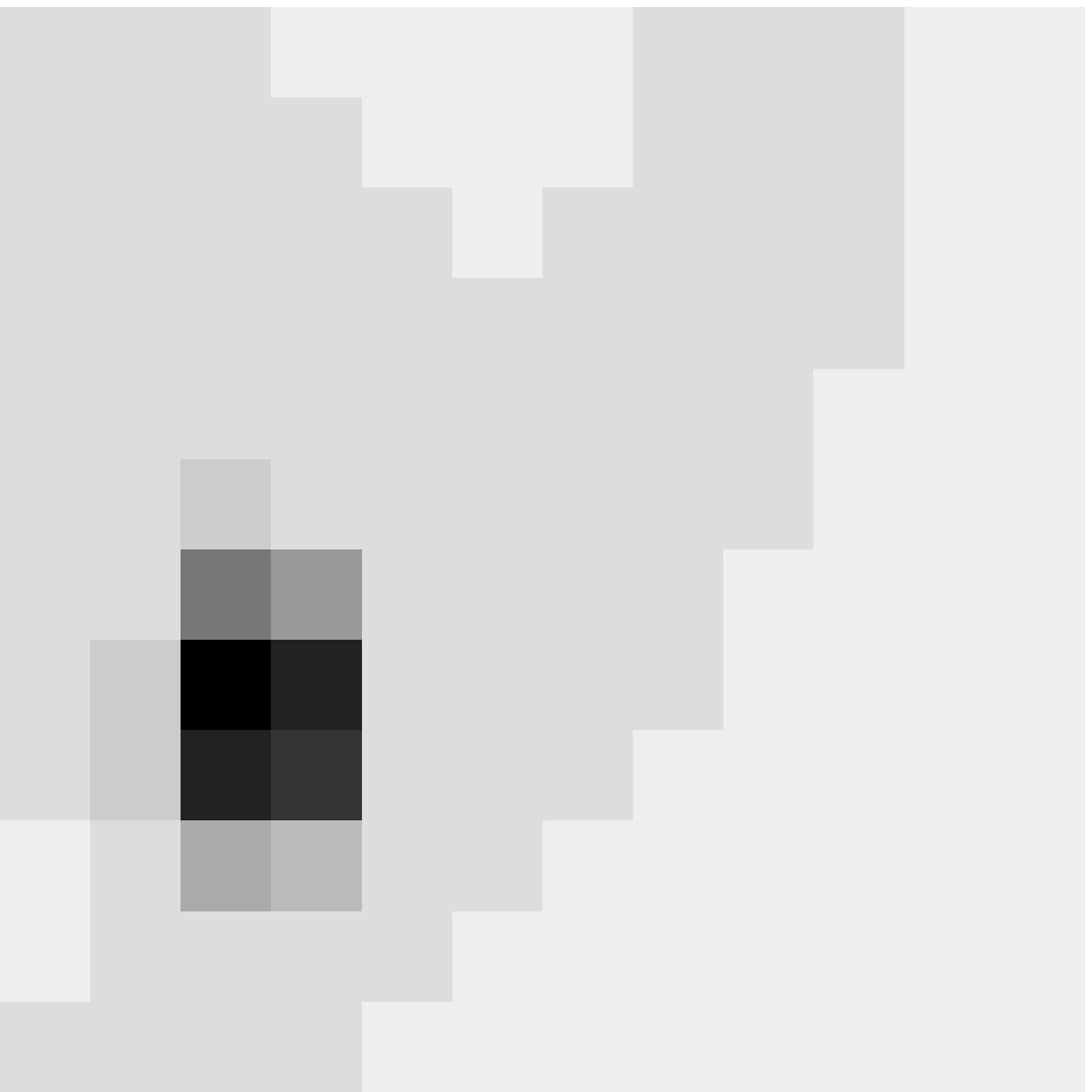}  \includegraphics[width=0.45\textwidth]{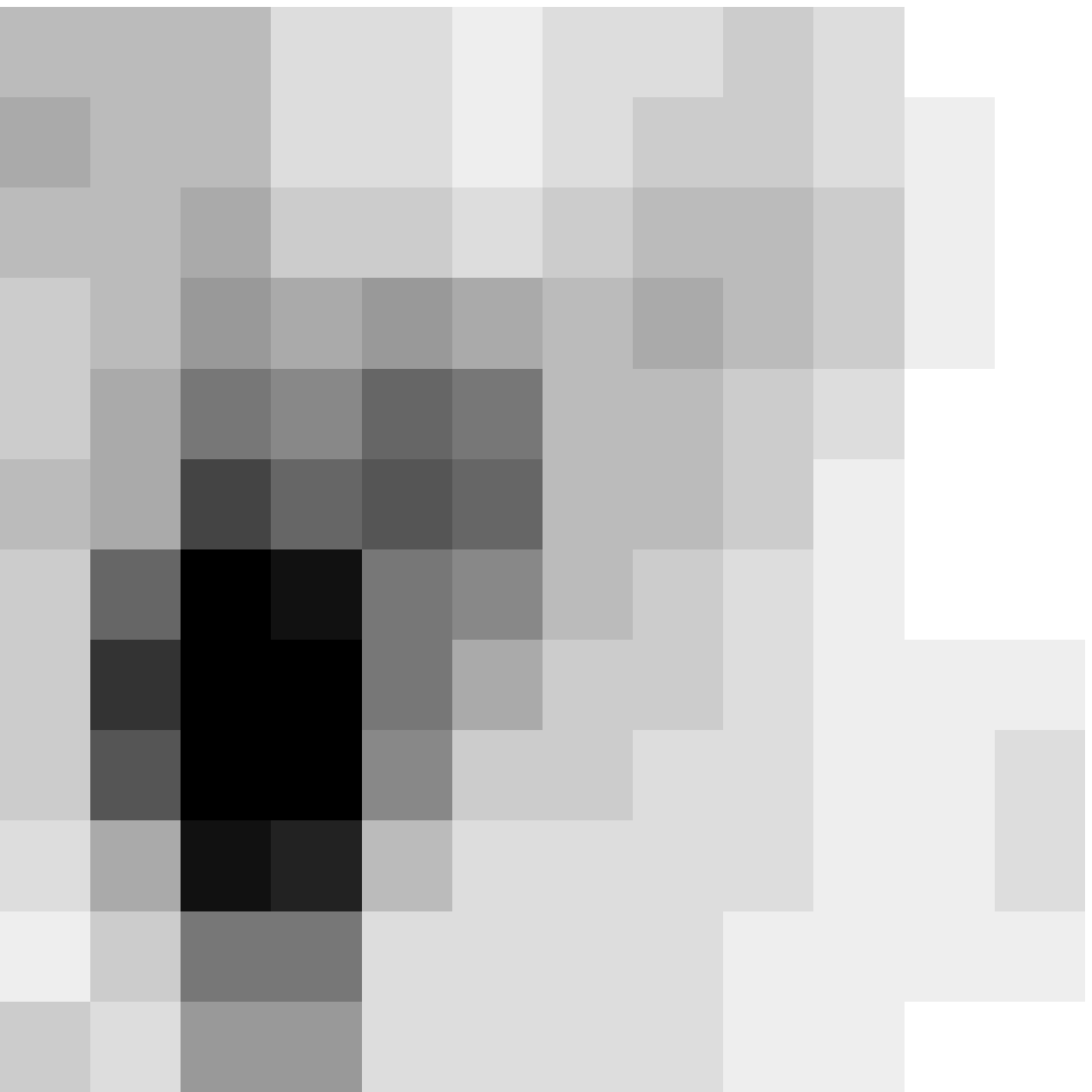} 
 \caption{Contrast-reversed images of the Sun at $2.40 \mu$m though the VIMS solar port -- both unstretched (left) and stretched (right) by displaying the square root of the DN value of each pixel.  The greyscale is such that 0 DN is 'white', and the peak solar signal is 'black'. To first order, the diffuse background is flat, but when stretched, the nonuniform features become clear.}
 \label{fig:image}
\end{figure}

\begin{figure} [htbp]
 \centering 
 \includegraphics[width=1.0\textwidth]{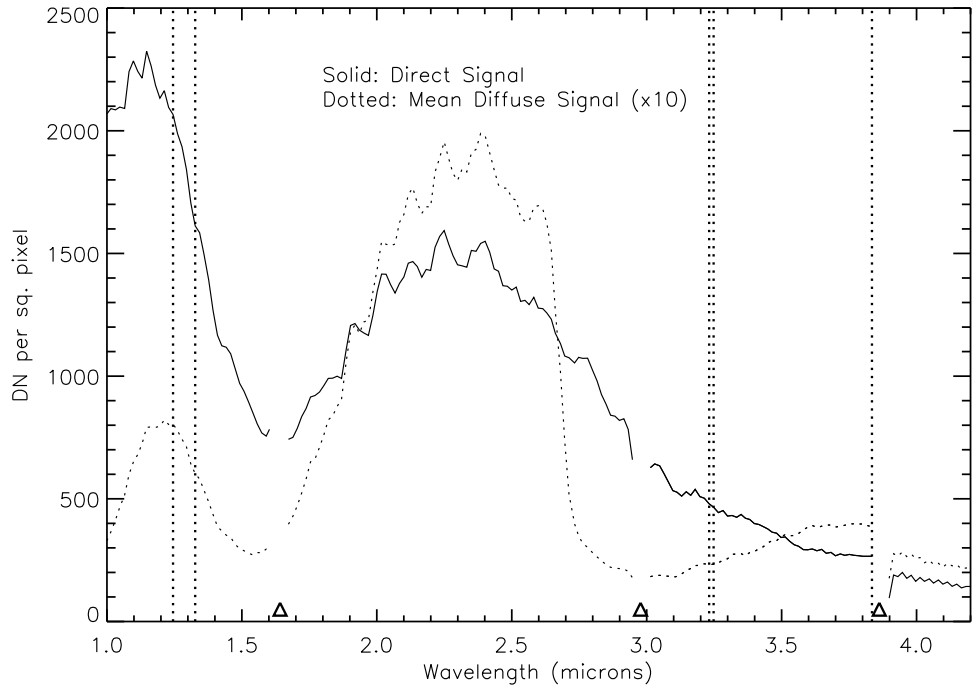}  
 \caption{Plot of the peak direct recorded signal (solid) and mean diffuse signal (dotted, magnified by 10 times) per pixel in the images taken of the Sun outside the rings on the Rev. 55 occultation. Peak values were measured by a Gaussian fit, and were recorded in units of DN per pixel.  Note that the signals have different spectral shapes, and that, in a 12 by 12 VIMS cube, the total diffuse signal is about an order of magnitude larger than the direct signal.  Triangles mark the locations of the edges of VIMS's order-sorting filters (which ensure only the listed wavelengths of light are measured by rejecting higher order signals from the diffraction grating), where the data become unreliable, while the vertical dotted lines mark spectral channels known for increased noise in calibration images.}
 \label{fig:dnsignal}
\end{figure}

The VIMS solar port is designed to attenuate the Sun enough to make it safe to observe with the VIMS instrument.  However, the optics that do this also produce abundant stray light within the instrument.  As a result, in addition to the normal solar image that can be fit to a two-dimensional Gaussian point-spread function (PSF), there is also a diffuse component that extends at least 6 solar diameters from the Sun (Figure \ref{fig:image}).  To first order, this diffuse component is flat over the 12 by 12 pixel images. At approximately 1/10th of the peak of the solar signal, the diffuse signal is $\sim10$ times larger than the flux within the nominal solar image when integrated over the entire cube (Figure \ref{fig:dnsignal}). In addition, the diffuse component is spectrally different from the direct component, being distinctly `redder'.   This greatly complicates any attempt to look for scattered light from millimeter-sized ring particles, but a method to exploit the stray light will be discussed in Section \ref{sec:missingtheory}. 

\subsection{Data Selection}

Of the eleven solar occultations taken before equinox in 2009 and observed by VIMS, nine cover the A ring, and six extend into the C ring.  The A ring occultations (Table \ref{table:aringobserv}) are mixed between nearly-radial occultations for which the Sun passed behind all of the rings (and then behind Saturn itself), and chord occultations for which the Sun passed behind one of the ansae, giving two `cuts' across the outer rings.  For the A ring, both the radial and chord occultations sample nearly uniformly in the radial direction as well as sampling only a limited range of longitude ($\lesssim 5^\circ$).  

The six occultations clearly covering the C Ring (Table \ref{table:cringobserv}) are also a mix of chord occultations and radial occultations.  As all of the chord occultations 'turn around' in the C ring, the data here have variable radial sampling, with the inner parts of the occultation (near the turnaround point) sampled more than outer parts.  

The criteria considered when deciding which data sets to use include the opening angle of the rings and the number of cubes within each ring.  For the A ring, occultations done later in the mission are almost opaque due to the low opening angle of the rings.  The C ring has the opposite problem; the large opening angles at the beginning of the mission meant that most of the sunlight is transmitted without interacting with the ring at all.  

\begin{table}
\begin{tabular}{|*6{l|}}
\hline
\multirow{2}{*}{Rev.}&\multirow{2}{*}{Date}&Ring Open.&Ave. &Number&Ave. \\
&&Angle ($^\circ$)&Long. ($^\circ$)&of Cubes&Trans.\\
\hline
09&08 Jun. 2005 (R)&21.45&78&47&0.348\\
43&24 Apr. 2007(E)*&12.77&297&70&0.270\\
55&03 Jan. 2008 (I)& 9.00&61&37&0.201\\
55&03 Jan. 2008 (E)& 9.00&143&37&0.041\\
59&20 Feb. 2008 (R)& 8.27&108&18&0.109\\ 
62&23 Mar. 2008 (I)& 7.79&41&15&0.099\\
62&23 Mar. 2008 (E)& 7.79&150&17&0.025\\
65&20 Apr. 2008 (I)& 7.36&45&16&0.099\\
65&20 Apr. 2008 (E)& 7.36&143&18&0.019\\
66&30 Apr. 2008 (I)& 7.21&46&16&0.098\\
66&30 Apr. 2008 (E)& 7.21&142&18&0.023\\
85&17 Sept. 2008 (I)&5.05&47&7&0.044\\
85&17 Sept. 2008 (E)&5.05&136&8&0.011\\
90& 24 Oct. 2008 (I)&4.49&50&7&0.037\\
90& 24 Oct. 2008 (E)&4.49&133&7&0.008\\
\hline
\end{tabular}
\caption{Observations of solar occultations covering the A ring.  Included is the date, the opening angle of the rings relative to the Sun at the time of occultation, the average longitude ($\phi$) of the observed place in the ring plane (measured relative to the sun-planet line), the number of cubes that clearly cover the A ring, and the average transmission measured.  Each occultation is marked as either a nearly-radial cut across the rings (R), or as the ingress (I) or egress (E) half of a chordal cut across the ring ansa. \\ $\ast$ The data set from the Rev. 43 occultation ingress starts near the inner edge of the A Ring, meaning the A Ring ingress was omitted from this table.}
\label{table:aringobserv}
\end{table}

\begin{table}
\begin{tabular}{|*6{l|}}
\hline
\multirow{2}{*}{Rev.}&\multirow{2}{*}{Date}&Ring Open.&Number&Ave.&Min.\\
&&Angle ($^\circ$)&of Cubes&Trans& Dist (Mm)\\
\hline
09&08 Jun. 2005 (R)&21.45&65&0.781&--\\
11&15 Jul. 2005 (R)&21.07&81&0.776&--\\
59&20 Feb. 2008 (R)& 8.27&51&0.498&--\\
62&23 Mar. 2008 (C)& 7.79&145&0.628&68.375\\
65&20 Apr. 2008 (C)&7.36&94&0.497&74.529\\
66&30 Apr. 2008 (C)& 7.21&84&0.520&83.844\\
\hline
\end{tabular}
\caption{Observations of solar occultations covering the C ring.  Included is the date, the opening angle of the rings relative to the Sun at the time of occultation, the number of cubes that clearly cover the C ring, and the average transmission measured.  Each occultation is marked as either a nearly-radial cut across the rings (R), or a chordal cut across the rings (C), in which case the minimum distance into the C ring that the chordal cut extend is listed in the last column.  Note that while the Rev. 62 and 65 chordal occultations cover most of the C Ring, the Rev. 66 chordal occultation only samples the outer half.}
\label{table:cringobserv}
\end{table}

\subsection{Transmission Spectra}
\label{sec:datatrans}

Transmission spectra of the main rings can be produced by summing the cubes over their spatial dimensions and normalizing to the solar spectrum as measured outside of the A ring.  This offers a high signal-to-noise spectrum of the ring's transmission properties in the near infrared, given the brightness of the Sun.  Combining repeated measurements at slightly different locations in the ring (sampled as the occultation progressed), we can increase signal-to-noise further at the expense of spatial resolution.  This gives a transmission spectrum with errors between 0.005 and 0.022 (in units of transmission).  

\begin{figure} [htbp]
 \centering 
 \includegraphics[ width=1.0\textwidth]{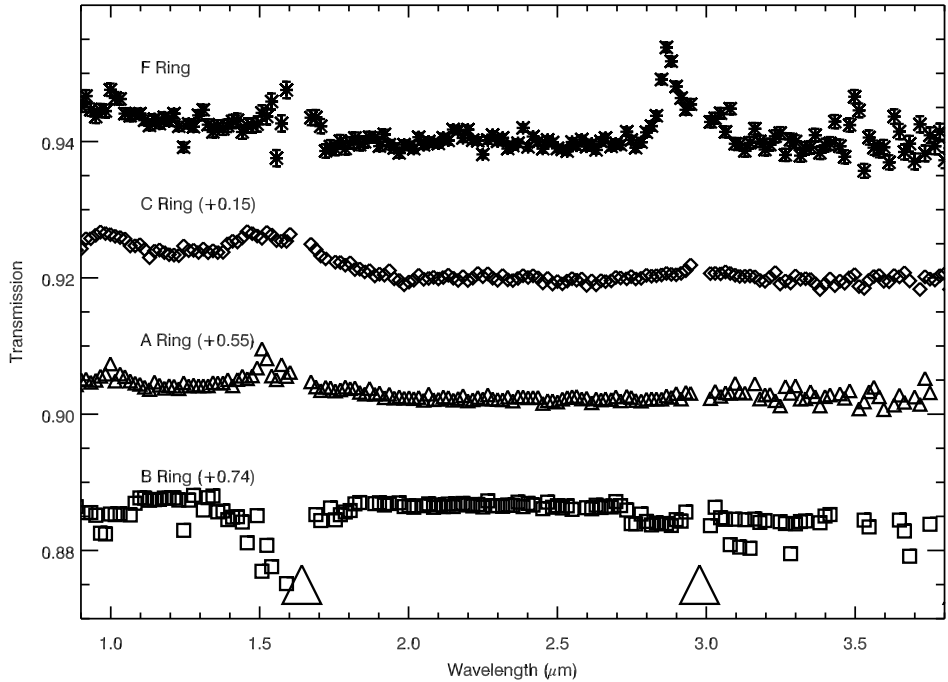} 
  \caption{Average transmission spectra of various regions of the rings as measured during the Rev. 9 solar occultation. Large triangles at the bottom of the plot mark the locations of VIMS's order-sorting filters (features at those locations are artifacts). Statistical error bars are not plotted for the A, B and C Ring spectra, as they are smaller than the plot symbol.  The A, B and C rings are also offset for clarity by the amounts indicated.}
 \label{fig:transspecrings} 
\end{figure}

In Figure \ref{fig:transspecrings}, we plot mean transmission spectra of the three main rings and the F Ring.  The spectra were constructed by fitting a gaussian curve to the image of the Sun in each wavelength, then taking the integral over that curve to find the total flux at that wavelength. Then an `average' spectrum for each area of the ring was produced by taking the mean over each cube `on' the rings, and normalizing to a solar spectrum obtained by taking the mean of cubes outside of the ring system.  

The main rings' transmission spectra show no obvious bands, and are remarkably flat in the region of 2 to 4 microns (the region from 4 to 5 microns is not plotted due to a much lower signal to noise ratio).   This is in marked contrast to the \textit{reflection} spectra of the main rings, which show strong water ice bands in this region (see \citet{2008Nicholson1} for a fuller discussion of the rings' reflectance spectra).  This indicates that the vast majority of ring particles are so large as to be opaque in the near infrared.  

However, not all regions of Saturn's rings behave in this matter.  Free ring particles in the tens of microns (or smaller) size range \textit{do} show prominent features in transmission, as is seen in our mean F Ring spectrum in Figure \ref{fig:transspecrings}, and described by \citet{2011Hedman1} in transmission spectra of the F Ring taken during stellar occultations.  Most visible in F Ring spectra is a strong increase in transmission at $\sim$ 2.9 $\mu$m due to the Christensen effect: the optical properties of water ice at this wavelength minimize absorption and internal reflection. \citep{2011Hedman1, 2011Vahidinia1}  

Other features, such as the peaks and dips near the order-sorting filters, are likely artifacts due to a lack of signal.  However, the slight `blue' slope around 1 to 1.5 microns may be a real measure of ring properties and will be discussed later in this paper.  

\section{Transmission Spectra Analysis}
\label{sec:datatransana}

\citet{2011Hedman1} introduce the ratio $\rho$ to measure the ratio in optical depth in and out of the 2.9 $\mu$m feature in stellar occultations.  In order to avoid contamination from reflected sunlight in addition to the transmitted starlight, they define $\rho$ as the ratio of optical depths at 2.9 and 3.2 $\mu$m, as the rings are dark in reflection at both wavelengths.  As solar occultations focus entirely on the dark sides of ring particles, the choice of a reference wavelength out of the 2.9 $\mu$m feature is less constrained.  We define $\rho_{2.5}$ as

\begin{equation}
\rho_{2.5} = \frac{\tau_{2.9}}{\tau_{2.5}},
\end{equation}

or the ratio between the optical depth of the 2.9 micron band (defined as the integrated optical depth from 2.82 to 2.93 $\mu$m) and the optical depth at 2.5 microns (defined as the integrated optical depth from 2.45 to 2.56 $\mu$m), with optical depths found the conventional way, from the transmission, $T = \exp{\tau/\mu}$.  2.5 $\mu$m was chosen as a reference wavelength based on the high signal to noise in this region of the solar spectrum as measured by VIMS.  

\begin{figure} [htbp]
 \centering 
 \includegraphics[ width=1.0\textwidth]{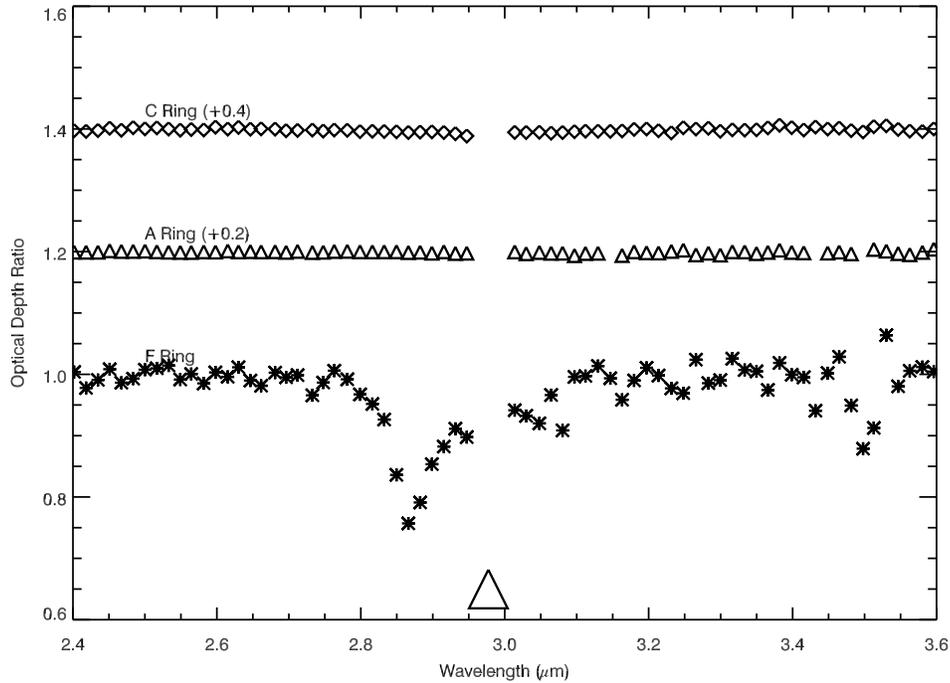} 
  \caption{The data from Figure \ref{fig:transspecrings}, replotted in units of optical depth and normalized so that $\tau$ at 2.5 $\mu$m is unity.  The F Ring (stars) shows a marked decrease in optical depth at 2.9 $\mu$m due to the presence of free-floating water-ice grains tens of microns in size.  The A (triangles) and C (diamonds) Rings show no such feature at 2.9 $\mu$m, limiting the number of free-floating ring particles smaller than 100 $\mu$m.  The region around 2.95$\mu$m, marked by the large triangle at the plot's bottom, was not plotted due to the presence of one of VIMS's order-sorting filters, as mentioned in Figure \ref{fig:dnsignal}.  The A, B and C rings are also offset for clarity by the amounts indicated.}
 \label{fig:transspecdetail} 
\end{figure}

Figure \ref{fig:transspecdetail} plots the composite spectra of the A, C and F rings from the Rev. 09 solar occultation in terms of the optical depth normalized to the optical depth at 2.5 $\mu$m.  In Figure \ref{fig:transspecdetail}, the 2.9 $\mu$m peak in the F Ring transmission spectrum is seen as a dip, while the A and C ring spectra continue to appear flat.   The measurements of $\rho_{2.5}$from six solar occultations (Revs. 09, 43, 55, 59, 62 and 65) are included in Table \ref{table:rho}.  From the table, the F ring shows a $\rho_{2.5}$ of between 0.77 and 0.86, with a mean value of 0.82 $\pm$ 0.03.  The A and C rings, however, yield values consistent with unity.  

%Enlarge Plot

%Enlarge Plot

\begin{table}
\begin{tabular}{|*2{l|}|*2{l|}}
\hline
Rev. & $\rho_{2.5}$&Rev. & $\rho_{2.5}$\\
\hline
\multicolumn{2}{|c|}{\textbf{F Ring}}&\multicolumn{2}{|c|}{\textbf{A Ring}}\\
\hline
9& 0.852 $\pm$ 0.004&9&0.9991 $\pm$ 0.0005\\
43&0.858 $\pm$ 0.002&43&0.9989 $\pm$ 0.0002\\
55 I&0.810 $\pm$ 0.008&55 I&0.9977 $\pm$ 0.0003\\
55 E&0.848 $\pm$ 0.008&65 I & 1.0177 $\pm$ 0.0003\\
\cline{3-4}
59& 0.809 $\pm$ 0.017&\textbf{Mean $\rho_A$} = & 1.003 $\pm$ 0.010\\
\cline{3-4}
62 I& 0.774 $\pm$ 0.009 &\multicolumn{2}{|c|}{\textbf{C Ring}}\\
\cline{3-4}
62 E& 0.782 $\pm$ 0.017 &9&0.9951 $\pm$ 0.0008\\
65 I& 0.820 $\pm$ 0.014 &59&1.0002 $\pm$ 0.0014\\
65 E& 0.824 $\pm$ 0.008 &62&0.9978 $\pm$ 0.0003\\
\cline{1-2}
\textbf{Mean $\rho_F$} = & 0.82 $\pm$ 0.03&65&1.0002 $\pm$ 0.0008\\
\hline
\multicolumn{2}{c|}{}&\textbf{Mean $\rho_C$} = & 0.998 $\pm$ 0.002\\
\cline{3-4}
\end{tabular}
\caption{Measure of the optical depth ratios between 2.9 $\mu$m and 2.5$\mu$m, as described by $\rho_{2.5}$.  Dusty water-ice rings, such as the F Ring, show a decrease in optical depth at 2.9 $\mu$m, resulting in $\rho_{2.5} < 1$.  Errors in the mean values listed for $\rho_{2.5}$ are calculated by taking the standard deviation of the set of measurements.}
\label{table:rho}
\end{table}

If we assume the A and C rings are a mixture of F ring-like material, with a $\rho_{2.5}$ equal to the mean F ring value of  0.82, and `large ring particles' with a $\rho_{2.5}$ of 1, we can set a limit on the amount of dusty or F-ring-like material. From the measured values of $\rho_{2.5}$, we conclude that neither the A nor the C Ring shows a significant difference from a flat spectrum.  The A Ring can contain less than 5.5\% (1 $\sigma$) by cross sectional area of F-ring-like material, while the C Ring can contain less than 1.4\% of F-ring-like material. From this, we can infer that free-floating ice grains in the tens of microns size range, capable of producing the Christiansen effect\citep{2011Hedman1}, are quite rare within the main rings, unlike within the F Ring.

\section{Diffraction Theory}
\label{sec:theory}

\subsection{Introduction}
\label{sec:theoryintro}
While, in the previous sections, we excluded a significant population of particles smaller than 100 $\mu$m in the main rings, somewhat larger particles can produce observable effects by diffraction, while being opaque.  It is to observe this diffraction that the spatial data taken by VIMS become useful.  

\begin{figure} [htbp]
 \centering 
 \includegraphics[ width=1.0\textwidth]{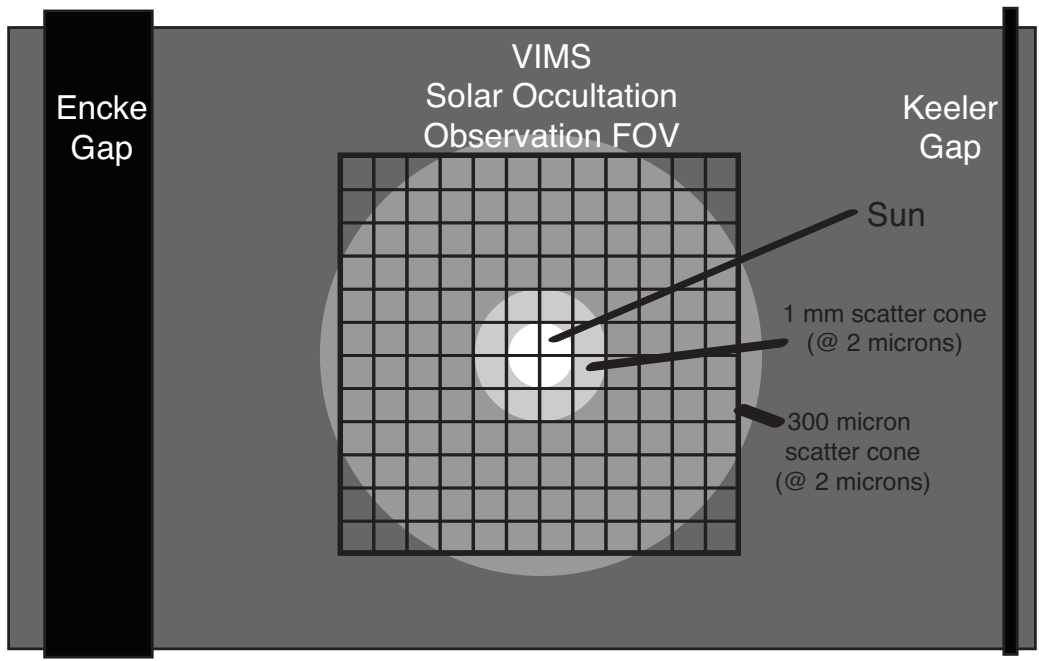} 
 \caption{Schematic diagram showing the size of a VIMS pixel, the 12 by 12 VIMS image taken during a solar occultation, and the Sun at Saturn during the 8 June 2005 solar occultation.  The estimated diffraction cones of a 1mm (light gray) and 300 $\mu$m (medium gray) ring particle at 2 microns are shown around the solar disk.  The Encke Gap (325 km) and Keeler Gap (40 km) at an appropriate Cassini-ring distance of 200,000 km are shown for scale.}
 \label{fig: ringdiagram}
\end{figure}

To first order, sunlight diffracted by ring particles of radius $a$ will scatter into a cone of angular radius $\theta \simeq \lambda/2a$.  Given VIMS's pixel size (0.5 milliradians) (see Figure \ref{fig: ringdiagram}), the solar diameter at Saturn ($\approx$ 1 milliradian) and operating wavelengths (1 - 5 $\mu$m), VIMS should be able to best image diffracted light from ring particles with a radius of several millimeters and less:

\begin{equation}
\label{eq:angscaling}
\theta_d \simeq \frac{\lambda}{2a} \simeq 1.0 \frac{\lambda/2 \mu\mathrm{m}}{a/1 \mathrm{mm}} \mathrm{mrad}.
\end{equation}

A full model of the diffraction of sunlight by ring particles will be presented in the following section.

\subsection{General Expression}
\label{sec:theoryintro}
The model of \citet{2000French1} was chosen as a simple representation of forward scattering and absorption in a ring.  \citet{2000French1} assume a simple truncated power-law size distribution and, for simplicity, neglect any contribution from multiple scattering -- which is a valid assumption for $\tau/2\mu \lesssim 1$.  We accept this for now, but in Section \ref{sec:multitheory}, we extend our analysis to include multiple-scattering for higher optical depths. As higher-order scattering broadens the phase function, ignoring higher-order effects will, in general, result in underestimates of the minimum particle size.  \citep{2000French1} 

This model states that the flux incident on the detector from light scattered by a uniform sheet of particles as a function of scattering angle, $F\left(\theta\right)$ is

\begin {equation}
\label{eq:modelbasic}
F\left(\theta\right) = F_0 \frac{\tau}{4 \pi \mu} e^{-\tau/\mu} \left<\varpi_0\right> \overline{P}\left(\theta\right) A
\end {equation}

where $F_0$ is the solar flux incident on the rings, $\mu$ is the cosine of the incidence angle, $\left<\varpi_0\right>$ is the single scattering albedo, assumed to be 0.5 for particles much larger than the wavelength of light being studied\footnotemark, $A$ is the solid angle of the detector (in this case, one VIMS pixel), and $\overline{P}\left(\theta\right)$ is the mean phase function of the diffracted light, normalized such that the integral over all solid angles is 4$\pi$ (thus the flux from scattered light integrated over all solid angles is $F_0 \tau/2\mu \exp\left(-\tau/\mu\right)$).  $\overline{P}\left(\theta\right)$ depends on the distribution of particle sizes assumed.  

Note that the optical depth, $\tau$, used in Equation \ref{eq:modelbasic} and for the rest of the paper (unless otherwise noted) is the extinction optical depth, which, for particles much larger than the wavelength of light, is twice that of the \textit{geometric} optical depth, $\tau = 2 \tau_{geo}$, where $\tau_{geo}$ is typically used in optical and near-infrared studies of the rings, including \citet{2000French1}.    

 \footnotetext{For our purposes, we may `lump' all light scattered at angles $\theta \gg \theta_d$ in with the light absorbed by the particles, so that the absorption coefficient, $Q_{abs} \approx 1$.  Since $Q_{ext} = Q_{abs} + Q_{sc} \approx 2$ for macroscopic particles, we have $\varpi_0 = Q_{sc}/Q_{ext} \approx 0.5$.  By the same token, we exclude all reflected light from the phase function, $\bar{P}\left(\theta\right)$. For further discussion on the importance of Q ,we refer the reader to \citet{1985Cuzzi1} and \citet{1987Roques1}.}
 
 For a full derivation of the model, please see \ref{append:theory}. We chose a truncated power law with particles between $a_{\min}$ and $a_{\max}$ in size, and with a power law index of $-q$.  To speed computational time over many orders of magnitude, we implement this in our code by two approximations valid over different angular regimes: the medium-angle case and the large-angle case, which are defined by the characteristic diffraction angle of the smallest particles in the size distribution, $\theta_2 = \lambda/2a_{\min}$.  These cases are also useful in understanding the behavior of the model.  
 
 The value of $\theta_2$ is unknown, because the minimum particle size is the quantity we are trying to measure. Given that the size of one VIMS pixel -- and coincidentally the solar radius at 9 AU -- is 0.5 milliradians on the sky, our data will be most sensitive to diffraction by particles with $x \lesssim 6000$, or, at 2 microns wavelength, particle sizes of 2 millimeters or less.  Barring a much-lower-than-expected minimum size cutoff, the large-angle scattering case will be most relevant, though we will include the medium-angle case in our calculations to account for the possibility of free-floating particles from $\sim$100 $\mu$m to $\sim$2 millimeters.  
 
 The large angle case, where the scattering angle, $\theta$ is much larger than the characteristic diffraction angle of the smallest particles ($\theta \gg \theta_2$), has a phase function of approximately 
 
 \begin{equation}
\label{eq:largebesselapprox_final}
\overline{P}\left(\theta\right) \approx \frac{4}{\pi\alpha}\left(\sin\theta\right)^{-3}\frac{x_{\min}^{2-q}}{q-2}.
\end{equation}

where the dimensionless size parameter $x_{\min} = 2 \pi a_{\min} / \lambda$ and $\alpha$ is a normalization factor given by 

\begin {equation}
\label{eq:alphadef}
\alpha=\left\{ \begin{array}{ll}
\ln \frac{a_{\max}}{a_{\min}} & q = 3\\
\frac{x_{\max}^{3-q} - x_{\min}^{3-q}}{3-q} & q \ne 3 \\
\end{array} \right.
\end{equation}

The medium angle case, where $\theta$ is in between the characteristic diffraction angle of the smallest and largest particles ($\theta_2 \gg \theta \gg \theta_1; \theta_1 = \lambda/2a_{\max}$), has a phase function of approximately 

\begin{equation}
\label{eq:intbesselapprox}
\overline{P}\left(\theta\right)\approx \frac{4}{\alpha}\left(\sin\theta\right)^{q-5}\mathcal{J}_0^\infty\left(q\right),\,\theta_1\le\theta\le\theta_2.
\end{equation}

The $\mathcal{J}_0^{\infty}(q)$ in Equation \ref{eq:intbesselapprox} is shorthand for $\int_0^{\infty}z^{2-q}{ J_1\left(z\right)}^2\,dz$.  It is nearly constant over the range of $2\le q \le 5$, except when $q$ approaches 2 or 5.  Previous studies indicate that $q$ is between 2.7 and 3.1 within the main rings, giving $J_0^{\infty}\approx0.5$ (\citealt{1985Zebker1}, \citealt{2000French1}, \citealt{2009Cuzzi1}).

At this point, we remind the reader that most of the light diffracted by the centimeter to meter-sized particles which dominate the main rings at near-infrared wavelengths is \textit{not} detectable, since it is confined to angles much less than the solar radius.  Our only hope is to detect the 'tail' of the scattering function, due primarily to millimeter and smaller sized particles, if they exist in sufficient numbers.  

Finally, we note that for the large-angle case, the \textit{slope} of $\bar{P}\left(\theta\right)$ is independent of $q$, but the absolute level depends on $q$ and $x_{\min}$ (or $a_{\min}$), while for the medium-angle case, the slope of $\bar{P}\left(\theta\right)$ depends on $q$, but the absolute level depends only weakly on $q$ (via $\mathcal{J}_0^{\infty}$) or $x_{\min}$ (via $\alpha$).  
\section{Spatial Data Analysis}
\subsection{Simple Attempts}
\label{sec:template}
Our goal is to detect and measure a faint 'halo' of diffracted light around the image of the occulted sun, in the presence of a much brighter background of instrument-scattered light and fit this by the method described in the previous section.  Our first approach was to attempt to create a template from the data of the unocculted Sun as seen through the solar port to serve as our comparison for cubes containing the occulted Sun.   We selected cubes outside of the F ring or inside the C ring to construct the template.  As observations were structured to give such windows on either side of solar occultations, these data were available for all occultations.  While the shape and spectrum of the diffuse background does vary depending on where in the field the Sun is, based on solar calibrations performed in flight, Cassini is a very stable platform for observations, and the movements of the Sun within the field during any single occultation are much smaller than a VIMS pixel.  As a result, the diffuse background changes little during an occultation, other than to scale with the total solar flux transmitted by the rings.  

Once we have a template, we can divide data cubes within the rings by that template.  Given the tiny levels of diffracted signal expected, data cubes from nearby radii in the rings were summed, creating composite data cubes for the average A Ring between the radii of 122,000 and 133,000 km and for the average C ring between 75,000 and 92,000 km.  Due to previous results which showed that the trans-Encke region of the A ring has a different particle size distribution than the middle and inner A ring \citep{2000French1, 1985Zebker1}, all A ring cubes outside of the Encke Gap were omitted from the average.   Cubes near ring edges were also omitted.  The number of cubes fitting these criteria from each occultation are listed in Tables \ref{table:aringobserv} and \ref{table:cringobserv} above.  

\begin{figure} [htbp]
 \centering 
 \includegraphics[ width=1.0\textwidth]{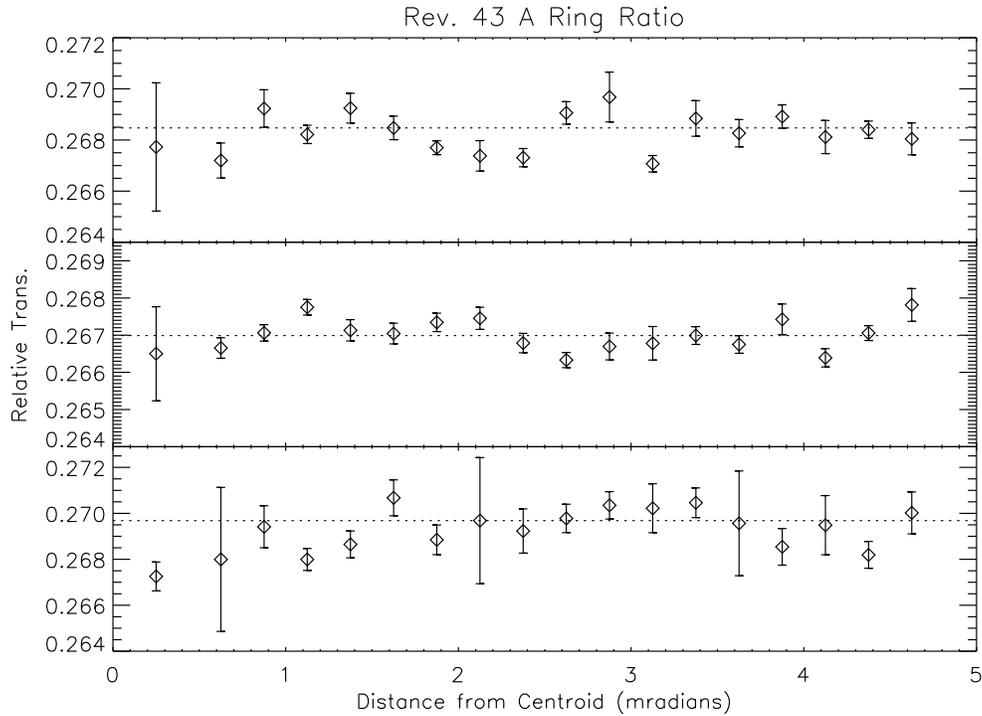} 
 \caption{A plot of the ratio of the composite A ring image from the Rev. 43 occultation to the template created from the same occultation versus  angular separation from the Sun.  Data are grouped in 0.25 milliradian bins, and the error bars mark one standard error of the mean for the binned data.  The dotted line is an average transmission for the area from 1 to 4 millradians from the Sun. Each panel is a different wavelength -- 1.2, 2.4 and 3.6 microns from top to bottom.}
 \label{fig:adiv}
\end{figure}

\begin{figure} [htbp]
 \centering 
 \includegraphics[ width=1.0\textwidth]{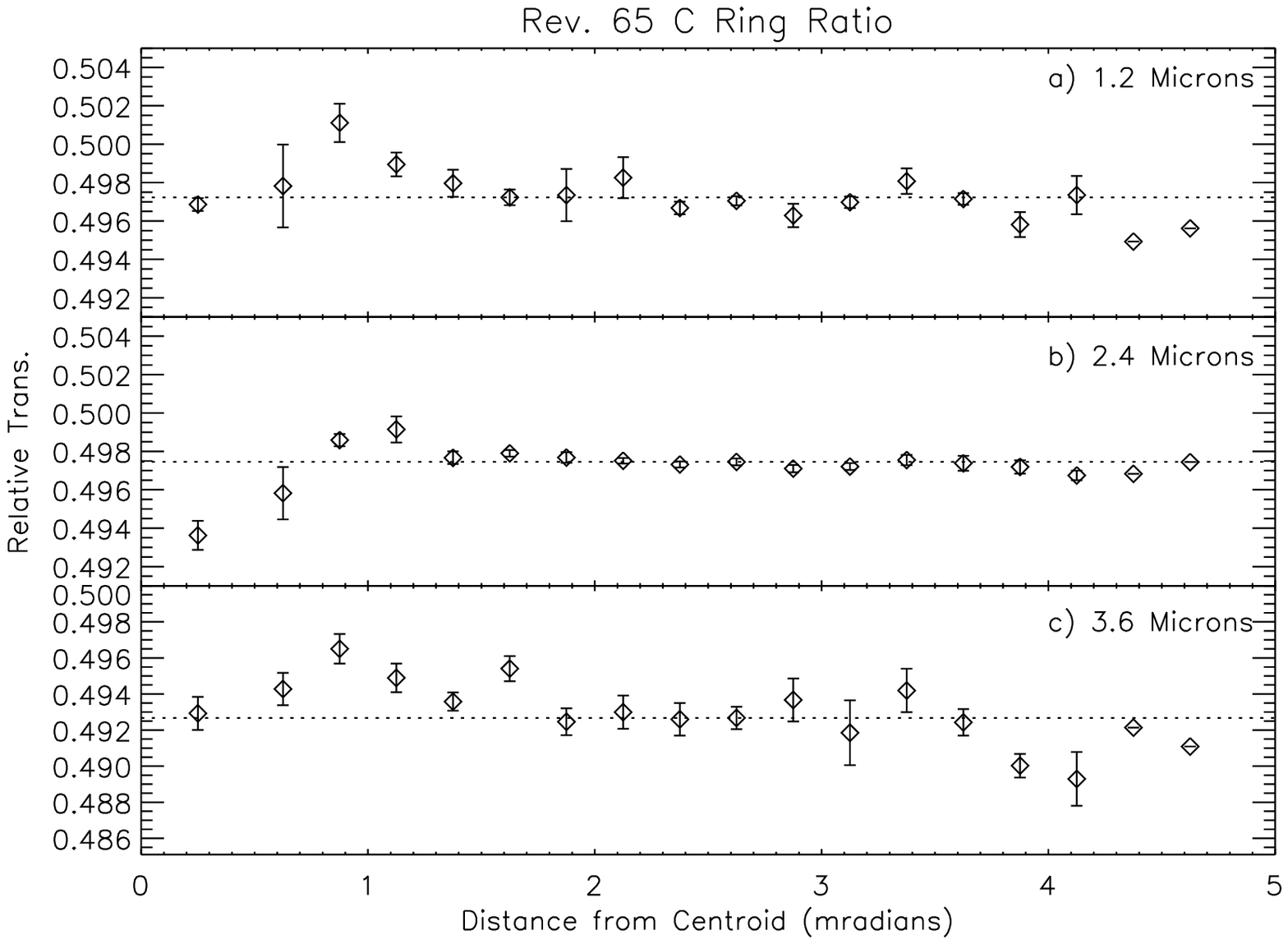} 
 \caption{A plot of the ratio of the composite C ring image from the Rev. 65 occultation to the template created from the same occultation versus  angular separation from the Sun.  Data are grouped in 0.25 milliradian bins, and the error bars mark one standard error of the mean for the binned data.  The dotted line is an average transmission for the area from 1 to 4 millradians from the Sun. Each panel is a different wavelength -- 1.2, 2.4 and 3.6 microns from top to bottom. Error bars are not plotted for the last two points, due to the paucity of data near the edge of the image. }
 \label{fig:cdiv}
\end{figure}

Figures \ref{fig:adiv} (the A ring, from Rev. 43) and \ref{fig:cdiv} (the C ring, from Rev. 65) show data from such ratio images, plotted in units of transmission (found by dividing the composite image by the templates constructed for each occultation using cubes containing an unocculted Sun).  The data from the ratio cubes were sorted by distance from the center of each pixel to the center of the Sun's image, and then binned in 0.25 milliradian (0.5 pixels) increments.  

%Enlarge Plot

%Enlarge Plot

Due to the non-zero instrumental background of diffusely-scattered light, transmission measurements can be recorded even far from the Sun itself, and are not themselves a sign of diffraction from ring particles.  Although at first glance the transmission profiles are `flat', closer inspection does show some evidence for diffracted light.  Figure \ref{fig:cdiv} shows a significant increase of a fraction of a percent in transmission at around 1 milliradian.   If a similar peak is present in Figure \ref{fig:adiv}, it is invisible compared to the pixel-to-pixel variation (as seen in the error bars, which mark the standard error).  Also note that the innermost datapoints (seen most clearly in Figure \ref{fig:adiv}.c and Figure \ref{fig:cdiv}.b, but present in others) show a significant \textit{decrease} in transmission relative to the 'far-field' mean transmission measured from 1 to 4 milliradians (and plotted as a horizontal line).  These pixels are within the 0.5 milliradians that define the solar angular radius as seen by VIMS, indicating that the transmission as measured by looking directly at the Sun is lower than that measured from the diffusely-scattered background, which is also produced by sunlight shining through the rings.  

This apparent contradiction can be reconciled by remembering that when sunlight is diffracted into a halo, it has to come from somewhere.  The individual pixels `on' the Sun will show some additional attenuation due to light scattered out of the beam.  However a wider-angle measurement -- like that of the stray light scattered within the VIMS instrument -- will collect both the direct and scattered light.  Were the Sun a point source and a VIMS pixel small enough to exclude all scattered light, the difference in transmission would correspond to a factor of two in optical depth.  Since neither is the case here, the difference is much more modest.  However, this difference in transmission can be measured, and, thus, can allow the amount of light diffracted at angles larger than a VIMS pixel to be measured, even if a clear diffraction halo is not seen (as in the A Ring measurements shown in Figure \ref{fig:adiv}).  This provides the concept behind our second approach, which we will elaborate on in the next section.

\subsection{Quantifying Transmission Differences}
\label{sec:missingtheory}
\begin{figure} [htbp]
 \centering 
 \includegraphics[ width=1.0\textwidth]{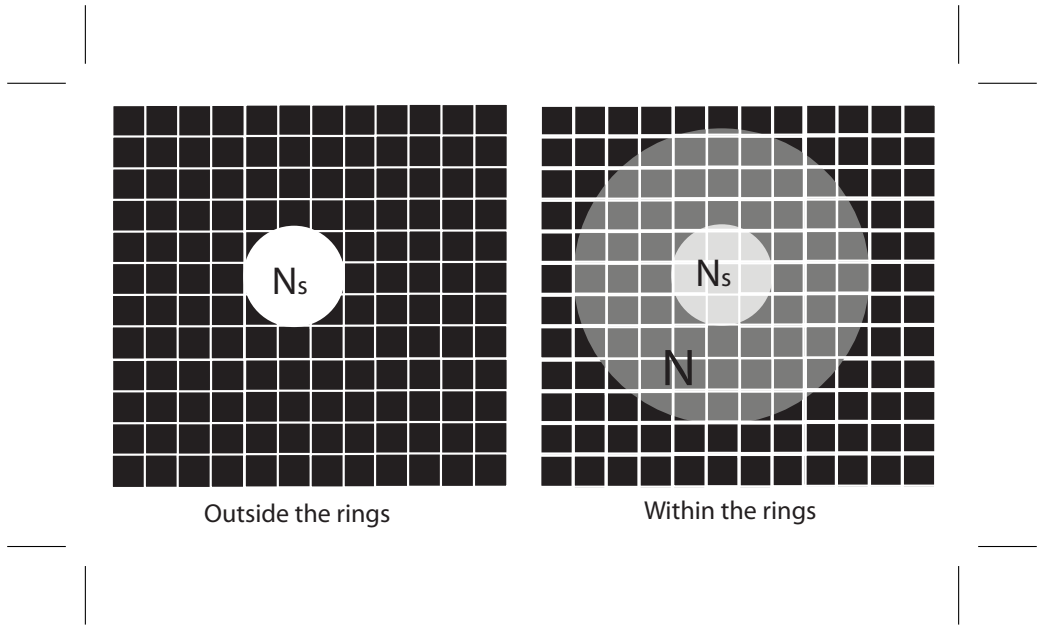} 
 \caption{Diagram showing our model for measuring light diffracted by the rings. The Sun takes up a small number of VIMS pixels, $N_s$.  While behind the rings, $N$ pixels (including the $N_s$ pixels) would show a small increase in flux from diffracted light.  If one were to coadd the image as a single measurement, the Sun would appear to have a higher transmission (and, thus, a lower optical depth) than if we were to only examine the $N_s$ pixels `on' the Sun.  This difference in optical depths should be easier to measure than attempting to measure the increase in signal in one (or a few) of the $N$ pixels, as it sums the entire effect of light scattered outside the central $N_s$ pixels.}
 \label{fig:ndiagram}
\end{figure}

Let us construct a simple model for imaging the Sun with VIMS.  The Sun is not a point source, so even an unocculted Sun will take up several VIMS pixels. For simplicity, we assume that the direct solar flux -- that not scattered by the solar port's optics or diffracted outside the sun's disc by small ring particles -- is confined to an area of $N_s$ pixels, which can be measured from the unocculted template we created in the previous section. When the Sun is behind the rings, there is an additional halo of diffracted light from small particles (defined as those capable of scattering light outside of the solar disc), covering $N$ pixels.   This is shown in Figure \ref{fig:ndiagram}.

Let the total direct (i.e., excluding that scattered within the VIMS solar port optics) unocculted solar signal be $S_s$, measured in DN per integration at a specific wavelength, $\lambda$.  In addition, as mentioned in Section \ref{sec:processinginstrument} and illustrated in Figure \ref{fig:image}, there is a diffusely-scattered background signal, spatially non-uniform, denoted by

\begin{equation}
 \sigma_{ob}\left(x,y\right) = \beta \left(x,y\right) S_s
 \end{equation}
 
 which we assume scales in brightness with the direct solar signal, but is spectrally different than the direct signal $S_s$ (Figure \ref{fig:dnsignal}).  The total integrated signal within the unocculted solar image (the $N_s$ pixels that are 'on' the Sun) can be written schematically as
 
 \begin{equation}
 \label{eq:unoccultsunsignal}
 S_0 \approx S_s + N_s \left< \sigma_{ob} \right>
 \end{equation}

where $\left< \sigma_{ob} \right>$ is a mean of $\sigma_{ob}\left(x,y\right)$.  We can estimate the values of $S_s$ and $\left< \sigma_{ob} \right>$ by fitting a two-dimensional Gaussian profile, plus a constant offset, to the central part of the template image.  

With the sun occulted by the rings, its total flux (direct plus diffracted) is reduced by a factor $T = e^{-\tau/2\mu}$.  The total flux from the sun is thus $TS_s$.  A portion of this flux has been diffracted by the rings at angles $\theta \lesssim \theta_2$.  We denote the fraction of the full solar flux diffracted into the range $\theta_s \le \theta \le \theta_2$ by $f$, where $\theta_s$ is the effective radius of the solar image in the VIMS cubes, or about 0.5 mrad.  Therefore, $N_s \approx \pi \theta_s^2$.  The diffracted flux is then $fS_s$, which is assumed to be spread uniformly over an area of $N \approx \pi \theta_2^2$ pixels, centered on the solar image.  

Three measured quantities are of interest in the cubes obtained during the occultation:

1. the background signal outside the diffraction halo,
\begin{equation}
\label{eq:backgroundocc}
\sigma_{rb} \left(x,y\right)= T S_s \beta  \left(x,y\right) = T  \sigma_{ob}\left(x,y\right) ;
\end{equation}

2. the background signal within the halo (i.e. in the annulus described by $\theta_s \le \theta \le \theta_2$), which has a mean value of 
\begin{equation}
\label{eq:rbprime}
\sigma_{rb^\prime} \left(x,y\right) = T\sigma_{ob}\left(x,y\right) + fS_s/N;
\end{equation}

3. the total signal within the solar image (defined as the same area of $N_s$ pixels above) of

\begin{equation}
S_r = \left(T-f\right) S_s+ N_s \left<\sigma_{rb^\prime}\right>
\end{equation}

or

\begin{equation}
\label{eq:occultsunsignal}
S_r = T\left(S_s + N_s \left< \sigma_{ob} \right>\right) - f \left(1- N_s/N\right) S_s .
\end{equation}

Given that we can measure $S_0$ and $\left< \sigma_{ob} \right>$ by fitting a two-dimensional Gaussian curve plus a constant to the data of the unocculted sun, as mentioned above, the only remaining unknown is $S_s$.  We can use Equation \ref{eq:unoccultsunsignal} to rewrite Equation \ref{eq:occultsunsignal} in terms of observables, rather than the unknown $S_s$, and we can normalize this by $S_0$ to get an effective transmission, $T_s$, measured only within the solar image:

\begin{equation}
\label{eq:occultsuntrans}
T_s = S_r/S_0 = T - f \left(1- N_s/N\right)\left(1 - N_s \left< \sigma_{ob} \right>/S_0\right).
\end{equation}

Note that all quantities in this expression, with the exception of $f$ (which is the measure of scattering by `small' particles in the ring, and which it is our goal to quantify), and $N$ (which is set by $\theta_2$ and thus the size of the smallest particles) can be directly measured from the data.  If we estimate a minimum particle size of $a_{\min} \approx 0.5$ mm, this gives $\theta_2\approx$ 2 mrad at 2 $\mu$m.  Since $\theta_s \approx 0.5$ mrad, the quantity $N_s/N \approx \left(\theta_s/\theta_2\right)^2$ is 1/16.  We can thus assume that $1-N_s/N \approx$ 1, and solve Equation \ref{eq:occultsuntrans} for $f$: 
 
\begin{equation}
\label{eq:scatterfracobs}
f  \approx \frac{T-T_s}{1 - N_s  \left< \sigma_{ob} \right>/S_0}
\end{equation}

T is most readily obtained from Equation \ref{eq:backgroundocc}, using the measurements of the instrument-scattered background.  In reality, $\sigma_{rb}\left(x,y\right)$ is spatially variable, so we use a Gaussian plus constant offset fit to occulted and unocculted cubes to find the local mean background in each image.  Then we obtain $T$ by dividing the constants of the two fits:

\begin{equation}
T =\frac{ \left<\sigma_{rb}\right>}{ \left<\sigma_{ob}\right>} .
\end{equation}

$T_s$ is also obtained by fitting offset gaussians to the occulted and unocculted solar images and integrating over the solar disk: 

\begin{equation}
T_s = \frac{\int F \left(x,y\right)\,dx\, dy}{\int F_0 \left(x,y\right)\,dx\,dy}.
\end{equation}

While gaussian curves are bounded at infinity, the constant background needed to properly fit the images are not.  We chose to assume the background under the solar image covers an area equivalent to the ellipse described by the fitted standard deviations of the gaussian function.  This `footprint' was chosen instead of a circle of angular radius $\theta_s$ (or the angular radius of the Sun at Saturn) to account for the distortion in the solar image: while the width of the gaussian in the $x$ direction matches the angular size of the Sun, the image appears stretched in the $z$ direction, as is clearly visible in Figure \ref{fig:image}.

Effectively what these calculations do is to estimate $f$ not from the diffracted light itself, but via its \textit{removal} from the direct solar flux.  The advantage of this somewhat indirect method is that the diffracted light is spread over $N$ pixels, while the solar image covers only $N_s$ pixels, where $N_s/N \ll 1$ for $a_{\min} \lesssim 0.5$ mm.  A secondary benefit is that the derived value of $f$ is almost independent of the unknown quality $N$, so long as $N_s/N \ll 1$.  Our first method (as explained in the previous section and shown in Figures \ref{fig:adiv} and \ref{fig:cdiv}) amounts to trying to measure the difference between the $\sigma_{rb}^\prime / \sigma_{rb} \approx  1 + f S_s/\left(N T \sigma_{ob}\right)$, which dwarfs the quantity of interest, $f$, by other factors, rather than measure it directly as we do here.  

\citet{2000French1} define a similar measure of the observed scattered light, $Q_{occ}$: the ratio of the observed optical depth, including some fraction of scattering, to the geometric (or absorption) optical depth, as defined in their equation 15; $\tau_{obs} = Q_{occ} \tau_{geo}$.  They define the total scattered flux measured in their equation 18, which can be written in our notation as 

\begin{equation}
f \approx \left(2-Q_{occ}\right) \frac{\tau}{2 \mu} e^{-\tau/\mu}
\end{equation}

As a test of concept, we can refer back to Figure \ref{fig:transspecrings}, which plots the direct solar signal (as measured by a Gaussian fit) in terms of transmission and as a function of wavelength.  We would expect that shorter-wavelength light would have less light scattered at angles large enough to be removed from the direct signal, producing a slightly blue slope as the redder regions of the spectrum had some light removed. In a qualitative sense, this can be seen in Figure \ref{fig:transspecrings}'s spectrum of the C Ring (and possibly the A Ring): the region of the spectrum blueward of $\sim$1.6 $\mu$m has slightly increased transmission than the rest of the spectrum. We also would expect that this effect would be somewhat dependent on optical depth -- at low or high optical depths, such a signal would not be as prominent as the intermediate optical depths that contain enough material to scatter, but not so much as to absorb the scattered light.  
\subsection{Measuring the Diffracted Light} 
\label{sec:specscatter}
Given the indications in Figures \ref{fig:adiv} and \ref{fig:cdiv} that $T_s$ is indeed slightly less than $T$, we can calculate $f$ numerically as described in Section \ref{sec:missingtheory}, by using a gaussian fit to both the template and individual cubes to calculate $T_s$ and $T$ -- and thus, $f$, the fraction of light diffracted out of the solar image.  As in the simple test performed above, it was necessary to take the mean of $f$ over the entire A or C ring -- with the same caveats of avoiding the trans-Encke region and the edges of the ring -- in order to achieve a satisfactory signal to noise level.  In addition, the data were binned by wavelength, taking the median of $f$ over 10 channels, with error bars calculated from the standard errors within each bin.  Based on those error bars, we focus on the region from 1.8 to 2.8 microns.    

Note that this bins data far more than in the simple plots we did in Section \ref{sec:template}. While Figures \ref{fig:adiv} and \ref{fig:cdiv} were means over wide ring regions, as are these measurements of $f$, here we bin ten adjacent wavelength channels as opposed to examining a single channel, and reduce an entire 144-pixel image into a single measurement (while in Figures \ref{fig:adiv} and \ref{fig:cdiv} each bin contains roughly a half-dozen points).  Thus, we should expect a corresponding reduction of noise.

In order to predict $f$ for a particular assumed size distribution, the model described in Section \ref{sec:theory} is used and integrated over an annulus centered on the Sun. A circle of radius 0.5 milliradians (1 pixel) was chosen as the inner boundary for the model's integral. However, as a result of the optics, the data show an clearly elliptical image of the Sun, and our measurements of $f$ (derived from the Gaussian fits to the data) take the apparent ellipticity into account. The distortion from the optics that produced an elliptical solar image could introduce a systematic difference between model and observation, but attempting to fit the image with a circular solar image would also introduce or exclude light.  Without a better mapping of the distortions caused by the boresight optics, an empirical measurement seems the best guess as defining the difference between `Sun' and `sky'.  

As the amount of scattered light drops off sharply with increasing angle, we assume an outer radius of infinity.  This introduces a negligible increase in the modeled value of $f$ for a given $a_{\min}$ compared to what we measure. Thus, $f$ is simply an integral over the intensity function, as specified in Equation \ref{eq:modelbasic}, divided by the unocculted solar flux: 

\begin {equation}
\label{eq:findf}
f = \int F\left(\theta\right) \,d \Omega /F_0 = \frac{\tau}{4 \pi \mu} e^{-\tau/\mu}\int_{\mathrm{0.5 mrad}}^{\infty}\int_0^{2\pi} \left<\varpi_0\right> \overline{P}\left(\theta\right)\,\theta d\phi d\theta
\end {equation}

\begin{figure} [htbp]
 \centering 
 \includegraphics[ width=1.0\textwidth]{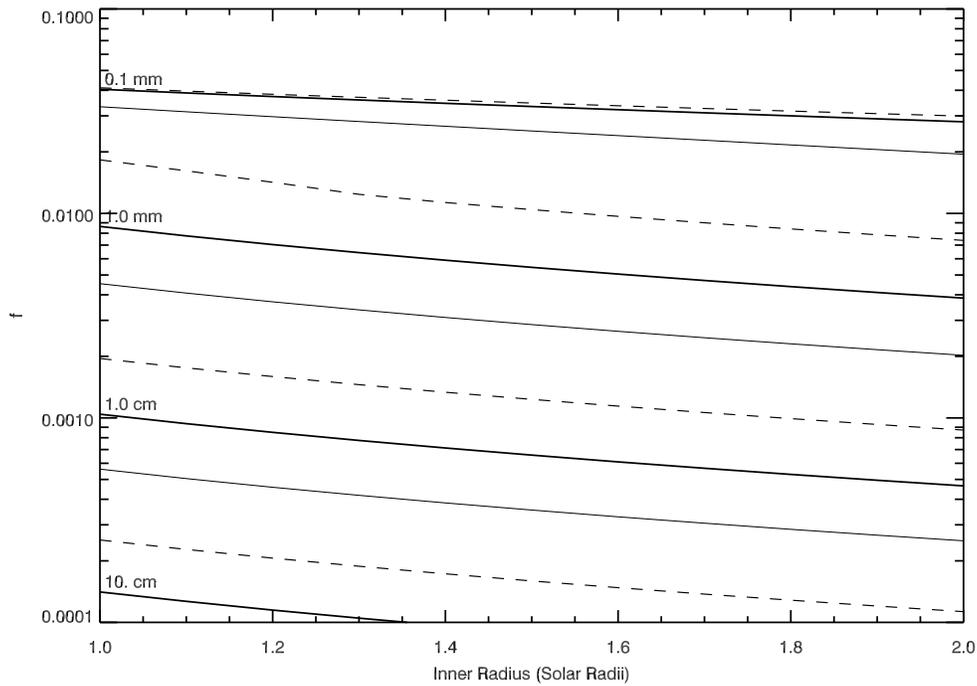} 
 \caption{Plots of the fraction of scattered light expected from hypothetical C ring models ($\tau/\mu = 0.5$, $q=3.1$, $a_{\max}=10$m, $a_{\min}$ as listed) versus the inner radius of the integral in terms of solar angular radius at Saturn.  While there is a clear dependence, varying the inner radius by a factor of two can, at most, produce an effect of a factor of two on inferred particle size.}
 \label{fig:thetadep}
\end{figure}

As a test of robustness, we integrated a hypothetical C ring model of $\tau/\mu = 0.5$, $q=3.1$, $a_{\max}=10$m and several lower particle size cutoffs for varying inner radii.  The results are shown in Figure \ref{fig:thetadep}.  Expanding the inner radius to an unphysical two times the solar angular radius in the image can reduce the minimum particle size by a factor of 2.  Consequently, any plausible error in estimating the `correct' annulus for the model would result in an overestimate of the particle size (as it seems unlikely that the most appropriate annulus would have an inner radius smaller than the solar radius).  

%Enlarge Plot

The C Ring occultations yield three data sets (those from Revs. 9, 62, and 65) which show a significant fraction of scattered light over the full spectral range considered (2 to 2.8 microns), and one more (Rev. 59) which shows a significant non-zero fraction of scattered light over part of this range.  Table \ref{table:cringobserv} includes the mean transmissions ($T$) and opening angles.  The Rev. 11 occultation does not give a significant detection; of the occultations, it has the highest background and it could be that statistical noise overwhelmed the signal.  

\begin{figure} [htbp]
 \centering 
 \includegraphics[ width=1.0\textwidth]{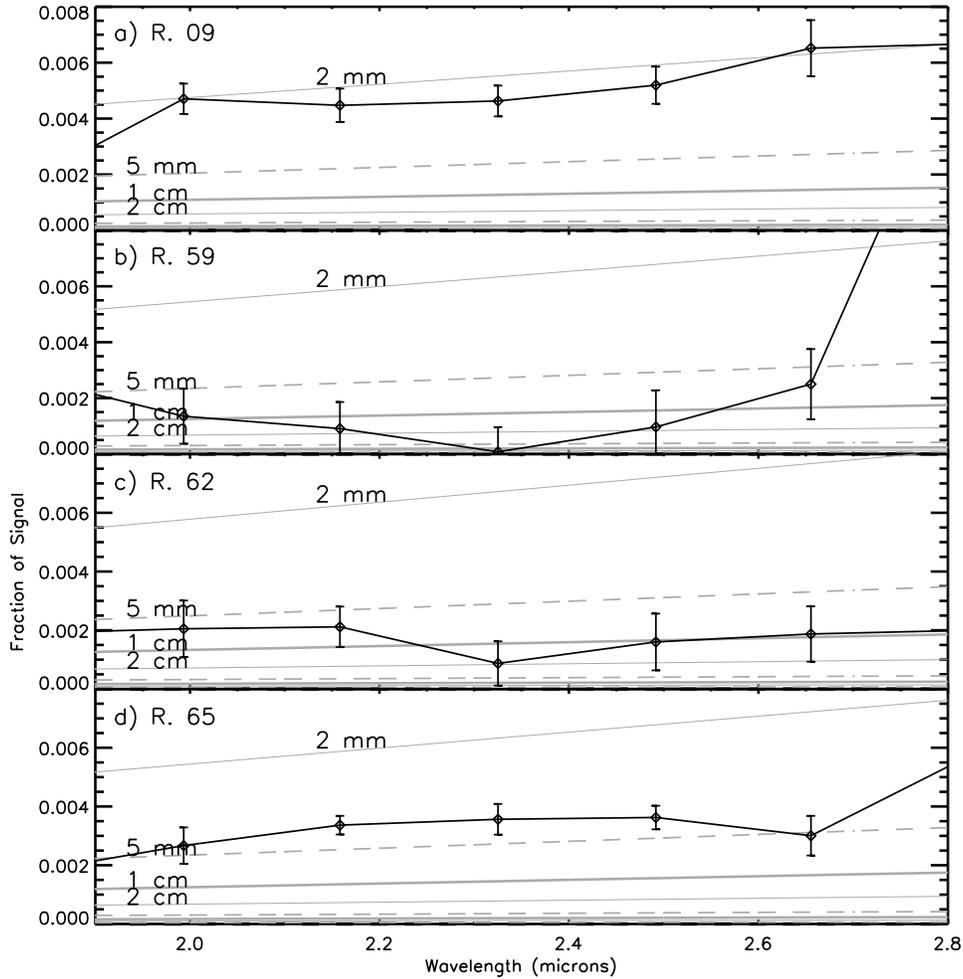} 
 \caption{Plots of the scattered light fraction, $f$, versus wavelength for four C ring occultations -- Rev. 9 (a), Rev. 59 (b), Rev. 62 (c), and Rev. 65 (d) -- calculated using Equation \ref{eq:scatterfracobs}.  The regularly-spaced arcs show models with $a_{\max} = 10$m, $q=3.1$, and minimum particle sizes of 2, 5, 10, 20, 50, and 100 mm (unlabeled). Note that Revs. 9, 62 and 65 show a significant fraction of scattered light that corresponds to a minimum particle size between 2 and 20 mm, while the Rev. 59 occultation only produces a marginal detection of diffracted light with a minimum particle size larger than 5 mm.}
 \label{fig:fractioncring}
\end{figure}

Figure \ref{fig:fractioncring} shows our three positive and one marginal detections.  As many of the C Ring solar occultations are non-uniformly sampled by radius, direct comparisons between occultations may be misleading if there are variations in particle size within the C Ring.  There are a mix of nearly radial occultations (Revs. 9 and 11), which sample all parts of the C Ring evenly, and occultations that cut across the ansae, which sample the innermost portions of the occultation more heavily (Revs. 59, 62, and 65). 

%Enlarge Plot

Given a model of the scattered light as discussed above and using values of $a_{\max}=$ 10 meters and $q=3.1 $\citep{1985Zebker1}, the three positive detections (Rev. 09, Rev 62 and Rev. 65 occultations) yield a minimum particle size of between 0.2 and 2 cm.  This range is of the same order as that derived by \citet{2008Marouf1} for the C Ring.  The slightly lower signal in the Rev. 65 and 62 occultations could indicate a slightly larger particle size cutoff in the inner portions of the C ring, as these two occultations oversample the inner regions, but the result is not at the 3$\sigma$ level given the error bars, especially those of the Rev. 62 occultation.  

The minimum particle size derived is somewhat dependent on the other model parameters $a_{\max}$ and $q$ -- a steeper power law or a smaller maximum particle size will increase the fraction of optical depth in smaller particles, and increase the amount of scattering at angles greater than $\theta_s$.  In our angular regime, namely that of large-angle scattering, the strongest effect is with $q$: a steeper power law (larger $q$) implies more particles with sizes small enough to scatter at the relevant angles.  Consequently, for a given value of $f$, a steeper power law leads to a larger $a_{\min}$.  $a_{\max}$ has only a weak effect; a larger maximum size reduces the number of particles per unit area for a given optical depth, slightly lowering the minimum size for the same value of $f$.  However, for $q>3$, as has been derived for the C ring \citep{1985Zebker1}, most of the cross-sectional area is in small particles, so a modest increase in the number of large particles produces an inconsequential effect on scattering at this angular scale.  

\begin{figure} [htbp]
 \centering 
 \includegraphics[ width=1.0\textwidth]{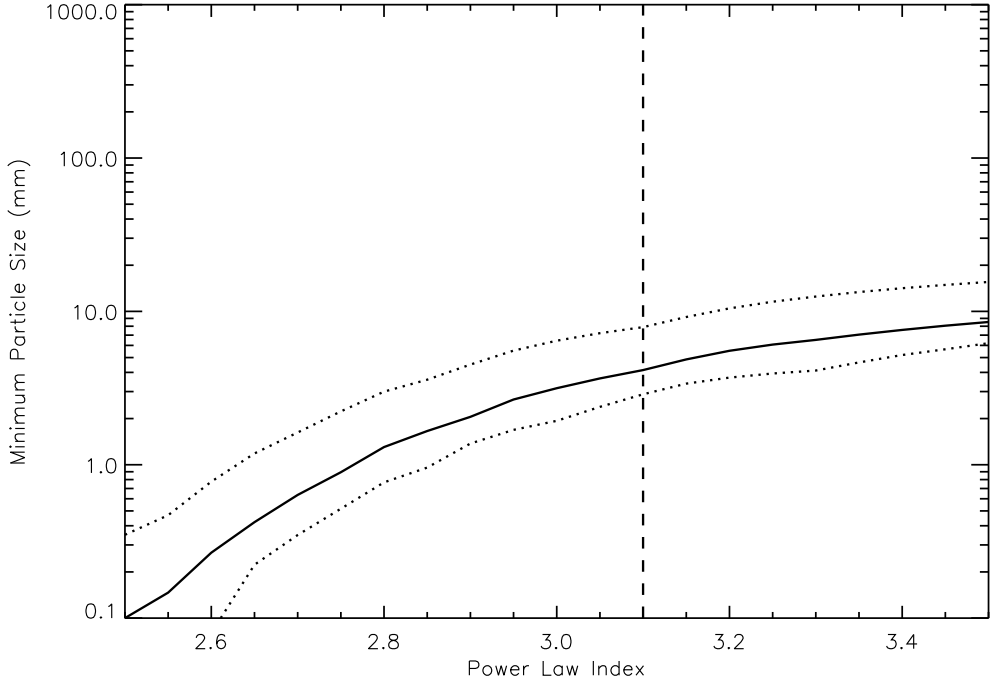} 
  \caption{A plot of $a_{\min}$ as a function of $q$ in the C ring, assuming a maximum particle size $a_{\max} = 10 m$ and for a wavelength of 2.3 $\mu$m. The dotted lines represent 1$\sigma$ errors on the estimates, combining both the differences between the calculated value of $a_{\min}$ from each occultation, and the errors of each occultation's $a_{\min}$ (calculated from the errors in $f$ calculated from binning nearby wavelengths).  The dashed line at $q=3.1$ represent previous estimates of the power law index for the C Ring. \citep{1985Zebker1, 2000French1, 2008Marouf1}}
 \label{fig:fractioncring2}
\end{figure}

Figure \ref{fig:fractioncring2} plots $a_{\min}$ vs. $q$ for the C ring.  The function was calculated by taking the scattering fraction from the Rev. 09, Rev. 62 and Rev. 65 occultations (the three in which a clear positive detection was made) at a wavelength of 2.3 $\mu$m, and calculating the $a_{\min}$ for a given $q$ needed to produce the observed scattered light.  The line plotted in Figure \ref{fig:fractioncring2} are then a mean of the values of $a_{\min}$ calculated from each of the three occultations.  At $q=3.1$, corresponding to previous estimates of the C ring power-law index \citep{1985Zebker1, 2000French1, 2008Marouf1},  we find a value of $a_{\min} = 4.1^{+3.8}_{-1.3}$ mm.  The C ring shows a robust value of $a_{\min}$ somewhere between 0.3 and 1 cm for values of $q$ between 2.95 and 3.5.  

%Enlarge Plot

\begin{figure} [htbp]
 \centering 
 \includegraphics[ width=1.0\textwidth]{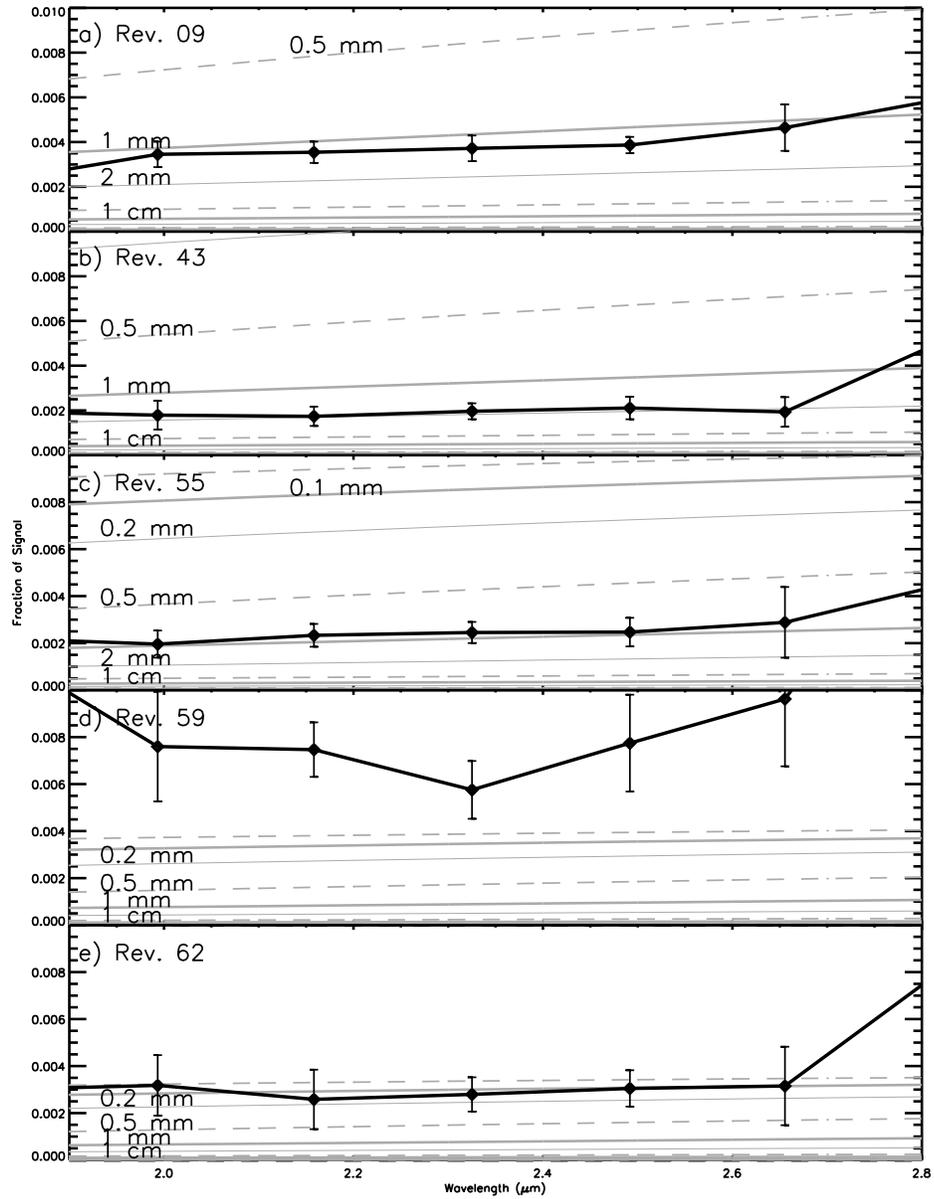} 
 \caption{Five A ring occultations -- Rev. 9 (a), Rev. 43 (b), Rev. 55 (c), Rev. 59 (d), and Rev. 62 (e) -- compared with single-scattering models ($a_{\max} = 10$m, $q=2.9$, and minimum particle sizes from 0.1 mm to 10 cm) with minimum particle size listed, calculated using Equation \ref{eq:scatterfracobs}.}
 \label{fig:fractionaringsingle}
\end{figure}

Completing the same analysis on the A ring -- shown in Figure \ref{fig:fractionaringsingle} --  shows a significant fraction of scattered light in five occultations (Revs. 9, 43, 55, 59, and 62) over the same wavelength range. Rev. 65 shows a partial detection over some of the range.  Note that comparing the far-field signal to the decrease in signal, and binning by wavelength, produces a far clearer detection in Rev. 43 than seen in Figure \ref{fig:adiv}.  However, unlike the C ring, which is homogenous and optically thin ($\tau/\mu \lesssim 1$), the A Ring is neither.  Those complicating factors, self-gravity wakes and the possibility of multiple scattering, are examined below and our simple model modified appropriately.

%Enlarge Plot

\section{The A Ring: Increased Optical Depth and Inhomogeneities}
\label{sec:wakes}
\subsection{Introduction to Self-Gravity Wakes}
\label{sec:introwakes}
Re-examining the assumptions made in the model described in Section \ref{sec:theory}, we see that one stands out.  The model assumes that the ring in question is made up of a thick slab with a homogenous distribution of particles.  However, the A ring is not well described by this model.  Observations  show that the A ring has an azimuthally-dependent optical depth, which varies by up to a factor of a few depending on the longitude relative to the planet-to-star direction that is sampled by the occultation \citep{2006Colwell1, 2007Hedman2}.  The accepted explanation for this variation, based on numerical simulations of this ring, is the presence of self-gravity wakes  \citep{1992Salo1}.

Because the self-gravity wakes are long aggregates of particles with a characteristic trailing orientation with respect to the radial direction, they change the optical depth depending on the cross section they present to the beam of light, which depends on the observed longitude with respect to the stellar direction, $\phi$.  They also don't show the simple 1/$\mu$ dependence of $\tau$ on opening angle, instead following a more complicated relation.  

Our next step in modeling ring scattering within the A ring is to account for the wakes within the ring.  

\subsection{Scattering with Opaque Wakes}
\label{sec:waketheory}
Following \citet{2007Hedman2}, we assume the A ring consists of a parallel series of cylindrical wakes of characteristic width $W$, height $H$, spacing $\lambda$ and alignment $\phi_W$ measured relative to the radial direction.  The wakes themselves are opaque, but the interwake `gaps' have a finite optical depth $\tau_G$ due to particles outside the wakes.  \citet{2010Tiscareno1} show that this is not an exact description of the wake behavior in dynamical simulations, but that this simple model reproduces optical depth measurements for opening angles larger than $\sim10^\circ$.  As none of the occultations we use for the A ring measurements are less than $\approx 8^\circ$, \citeauthor{2007Hedman2}'s wake model should be sufficient for our purposes.   

We assume that the particles within the gaps form a homogenous layer so that the same model used earlier applies within the gaps.  If $f_W$ is the fraction of ring covered by the wakes as viewed by VIMS (which depends both on the opening angle of the rings ($B$), and the longitude ($\phi$), as well as the parameters  $W$, $H$, $\lambda$ and $\phi_W$), then the fraction of scattered light can be written as 

\begin{equation}
\label{eq:wakescatter}
f = \left(1-f_W\left(B, \phi\right)\right)\frac{\tau_G}{4 \pi\sin B} e^{-\tau_G/\sin B} \int \left<\varpi_0\right> \overline{P}\left(\theta\right) \, d\Omega,
\end{equation}

where the integrand is calculated as before.  Not that for $f_W = 0$ (and $\tau_G$, the extinction optical depth of the interwake material, equal to $\tau$), this equation reduces to the simpler case used in the previous section.  

A full derivation for $f_W$ is given by \citet{2007Hedman2}, resulting in the expression

\begin{equation}
\label{eq:fwvalue}
f_W = \left|\frac{H \sin \left(\phi-\phi_W\right)}{\lambda \tan B}\right|\sqrt{1+\left[\frac{W \tan B}{H \sin \left(\phi-\phi_W\right)}\right]^2}.
\end{equation}

The values of $\phi$, the longitude of the area sampled, and $B$ are known for each occultation and are listed in Table \ref{table:aringobserv}.  $\phi$ changes slightly as the occultation progresses, as none are totally radial occultations, but for the regions of the A ring sampled, this change is small.  Repeated stellar occultations suggest that $\tau_G$ is between 0.3 and 0.6, $H/\lambda$ is between 0.09 and 0.12, and $W/\lambda$ is between 0.3 and 0.65  for the A ring\citep{2010Nicholson1}. Note that \citeauthor{2010Nicholson1}'s values for optical depth ($\tau_g$) correspond to absorption, so we have used the equation $\tau_G = 2 \tau_g$ to derive values for extinction optical depths within the gaps. These and other studies of A ring photometry show that the wakes are oriented to have a peak transmission at $\phi_W \approx 70^\circ$ and $250^\circ$ longitude (with 0$^\circ$ being the direction to the Sun (or star) from Saturn)\citep{2010Nicholson1}.

The effect of the wakes on $f$ is not simple.  While $f_W$ decreases the ring area which provides the scattered signal, replacing $\tau$ by the much smaller $\tau_G$ increases the amount of scattered light available when $\tau/\mu > 1$, which is usually the case in solar occultations by the A Ring.

\subsection{Effects of Multiple Scattering}
\label{sec:multitheory}

Our first model assumed that all light interacts with a ring particle once and is absorbed, (singly-)diffracted or transmitted.  However, in reality the ring particles we are considering are far smaller than the thickness of the ring, so multiple scattering is possible. For $\tau/2\mu \ll 1$, the contribution from light diffracted more than once is small.  However, even when we consider an expected normal optical depth of 0.3 to 0.65 (the estimated extinction optical depth between self-gravity wakes in \citet{2007Hedman2}), only the Rev. 9 occultation at $B=21.5^\circ$ (and the lowest optical depth estimate of the gap material) satisfies $\tau/2\mu < 1$.  

\citet{1985Zebker1}, in analyzing the low incidence Voyager radio occultations, developed a scheme for handling multiple scattering in a thin ring. They treat the ring as $N$ layers of optical depth $\tau_1 = \tau/N$, where $\tau_1 \ll0.5$, so that within each layer, the single scattering approximation holds.  This model allows for multiple scattering (to degree $N$) by calculating the fraction of absorption, scattering or transmission through each layer and treating it as a sum of terms to produce the intensity function. The phase function for multiple scattering is treated of a convolution of single scattering, as \citet{1985Zebker1} do in their Equation 7. In the notation used in this paper, we can write their equation as 

\begin{equation}
\label{eq:zebkerlayers}
\frac{I_{sca}\left(\theta\right)}{F_0} = \sum_{k=1}^N \left(\begin{array}{c}N\\k\end{array}\right)e^{-\tau \left(N-k\right)/N \mu} \left[\frac{I_{1}\left(\theta\right)}{F_0}\right]^k,
\end{equation}

where $I_1\left(\theta\right)$ is the intensity distribution from single scattering within a layer of ring, as calculated from Equation \ref{eq:modelbasic} (but without the solid angle that changes an intensity into a flux), using $\tau_1$ as the optical depth.  $\left[I_1\right]^k$, represents the $k$th convolution of $I_1$ with itself, so each term of the sum represents the contribution of $k$th order scattering to the whole, with an attenuation factor to account for absorption by the $N-k$ other layers, and a combinatoric factor to account for the $k$ layers chosen from $N$ to scatter photons.  We can also re-write Equation \ref{eq:zebkerlayers} in terms of the phase function for single scattering, $P\left(\theta\right)$, to remove quantities not dependent on $\theta$, and to bring out the `hidden' $\tau_1$ in the intensity distribution: 

\begin{equation}
\label{eq:phaselayers}
\frac{I_{sca}\left(\theta\right)}{I_0} = e^{-\tau/\mu}\sum_{k=1}^N \left(\begin{array}{c}N\\k\end{array}\right)\left(\frac{\varpi_0\tau}{4 \pi N \mu}\right)^k \left[P\left(\theta\right)\right]^k .
\end{equation}

$N$ is an approximation for the number of particles thick the ring is.  As mentioned earlier, the Voyager radio occultation was most sensitive to suprameter particles, with smaller particles sensed only as a differential optical depth between the two wavelengths of radio waves transmitted through the rings.  As the rings are thin relative to meter-sized particles, even when considering the slant-path at low incidence angles, \citeauthor{1985Zebker1} could assume $N$ was small and searched for the value of $N$ which best agreed with the data. However, in the case of millimeter-sized particles, the rings are no longer physically thin relative to the particle diameter, even at normal incidence angles.  Thus, rather than $N$ being a few, it becomes on the order of a thousand.  

If we let $N$ become large, then the equation becomes 
\begin{equation}
\label{eq:sumseries}
\frac{I_{sca}\left(\theta\right)}{I_0} = e^{-\tau/\mu}\sum_{k=1}^\infty \frac{1}{k!}\left(\frac{\varpi_0\tau}{4 \pi \mu}\right)^k \left[\bar{P}\left(\theta\right)\right]^k .
\end{equation}

\begin{figure} [htbp]
 \centering 
 \includegraphics[ width=1.0\textwidth]{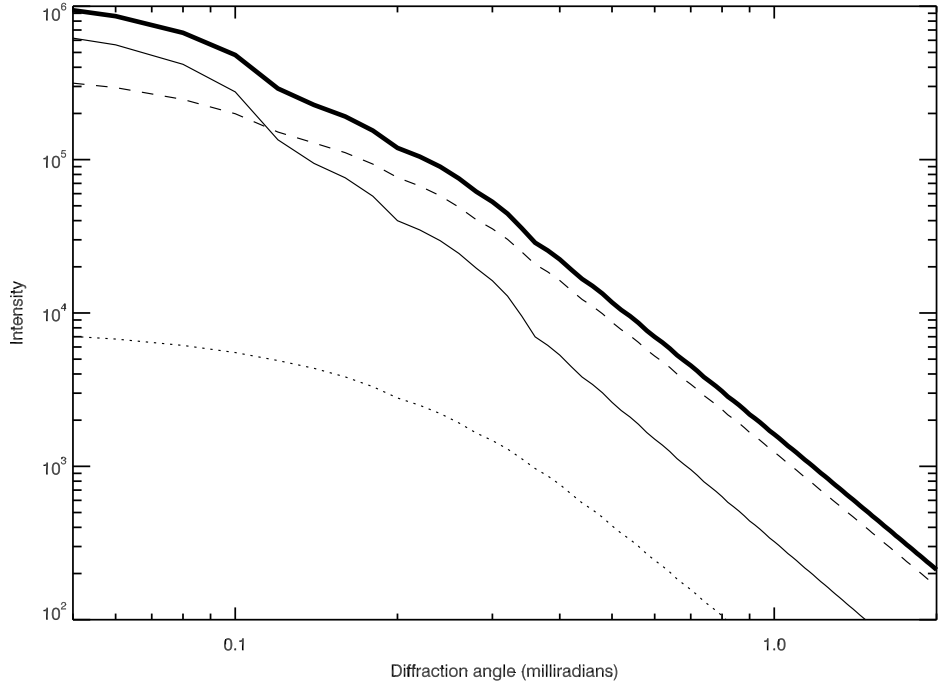} 
 \caption{Plot of the contributions of single (solid), double (dashed) and triple (dotted) particle scattering to the total intensity (thick) of the scattering versus diffraction angle for an optical depth of $\tau/\mu=1$, a wavelength of 2 $\mu$m, and a power-law particle-size distribution of index $q=2.9$, from 1 mm to 10 m. These conditions are roughly analogous to the A Ring.  Note that, in fact, double-particle scattering dominates over single-particle scattering at $\sim$1 milliradian where our observations are most sensitive. Triple-particle scattering and higher-order terms (not shown) make up a minor part of the scattering function.}
 \label{fig:iplot}
\end{figure}

We will be using this equation to include the effects of double- and triple-particle scattering.  Higher order terms are small relative to these terms, so were omitted.  From Figure \ref{fig:iplot}, we can see that double-particle scattering produces the dominant effect at angles larger than the $\sim$ 0.5 milliradians that marks the size of the solar image (and, thus, the minimum angle required to remove light from the signal), confirmation of the necessity of accounting for multiple-particle scattering. Conceptually, this can be explained as the more times a photon is scattered, the broader the diffraction cone becomes. If little light is being singly scattered at a certain angle, doubly scattered light will dominate if the ring is optically thick enough.  The decrease in intensity from single and double-particle scattering to triple-particle scattering justifies our neglect of higher-order terms.  

%Enlarge Plot

\subsection{Measuring Diffracted Light in the A Ring}
\label{sec:ameasure}
Now that we have discussed the complicating effects of multiple-order scattering and self-gravity wakes, we can add them to the model.  Note that the two effects to an extent work against each other: multiple-order scattering will increase the amount of scattered light for a given optical depth, while self-gravity wakes will lower the material available to scatter light, which will decrease the scattered light in general (as well as add a longitude-dependent term).  It is not obvious which (if either) effect will dominate at the scales we are interested in for this problem.  

To model the A Ring, we had to choose parameters to represent the self-gravity wakes.  The wake dimensions of $W/\lambda$=0.5 and $H/\lambda$=0.1 were chosen as representative parameters from the stellar occultation data discussed in Section \ref{sec:waketheory}.  Individual values of $\tau_G$ for each cube were calculated based on those numbers and assuming $T =\left(1- f_W \right)e^{-\tau_G/2\sin B}$, with $T$ being the calculated transmission in that cube and $f_W$ calculated from Equation \ref{eq:fwvalue}.  

\begin{figure} [htbp]
 \centering 
 \includegraphics[ width=1.0\textwidth]{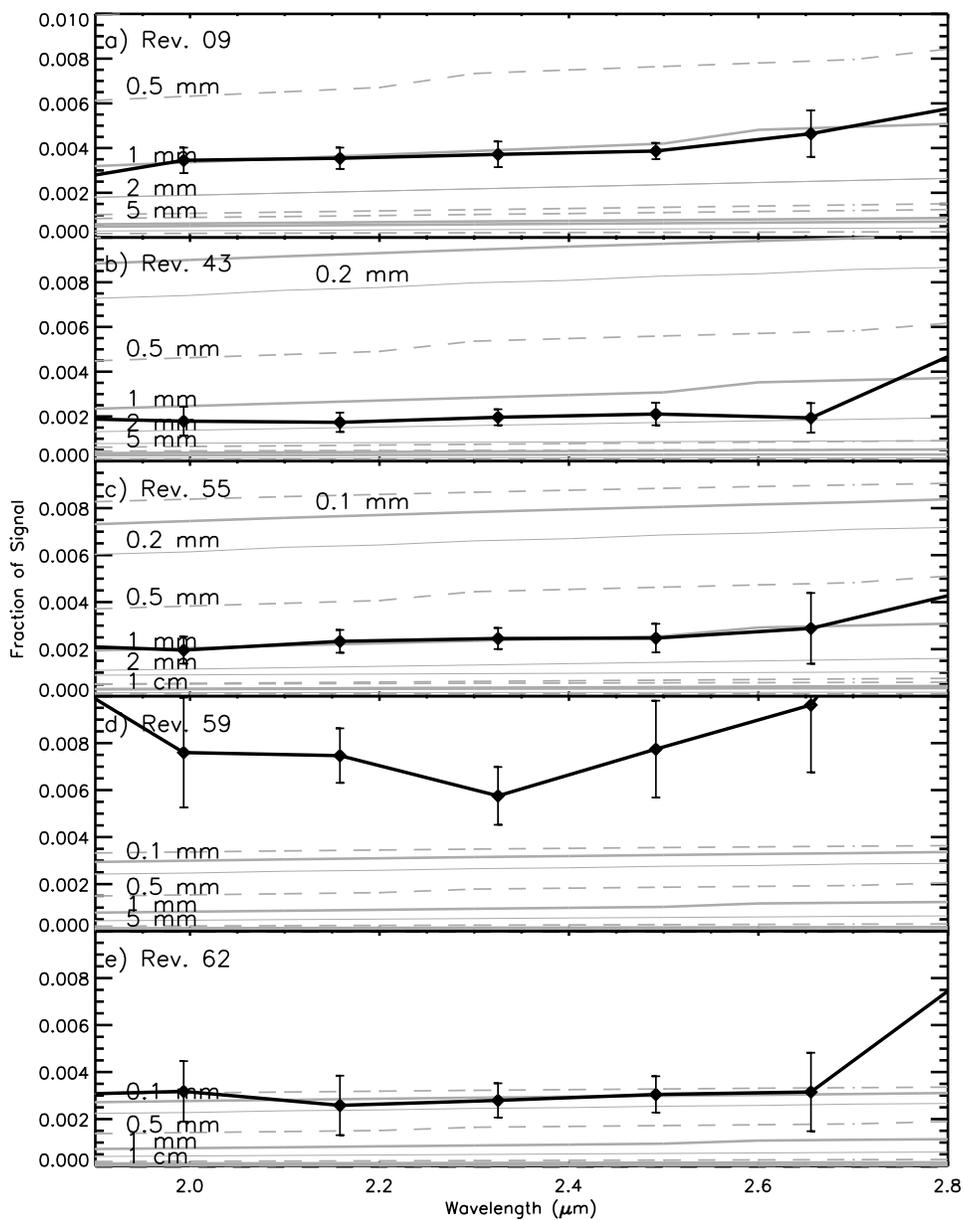} 
 \caption{Five A ring occultations -- Rev. 9 (a), Rev. 43 (b), Rev. 55 (c), Rev. 59 (d), and Rev. 62 (e) -- compared with models ($a_{\max} = 10$m, $q=2.9$, minimum particle sizes from 0.1 mm to 10 cm, self-gravity wakes and multiple scattering included) with minimum particle size listed. }
 \label{fig:fractionaringwwakes}
\end{figure}

\begin{figure} [htbp]
 \centering 
 \includegraphics[ width=1.0\textwidth]{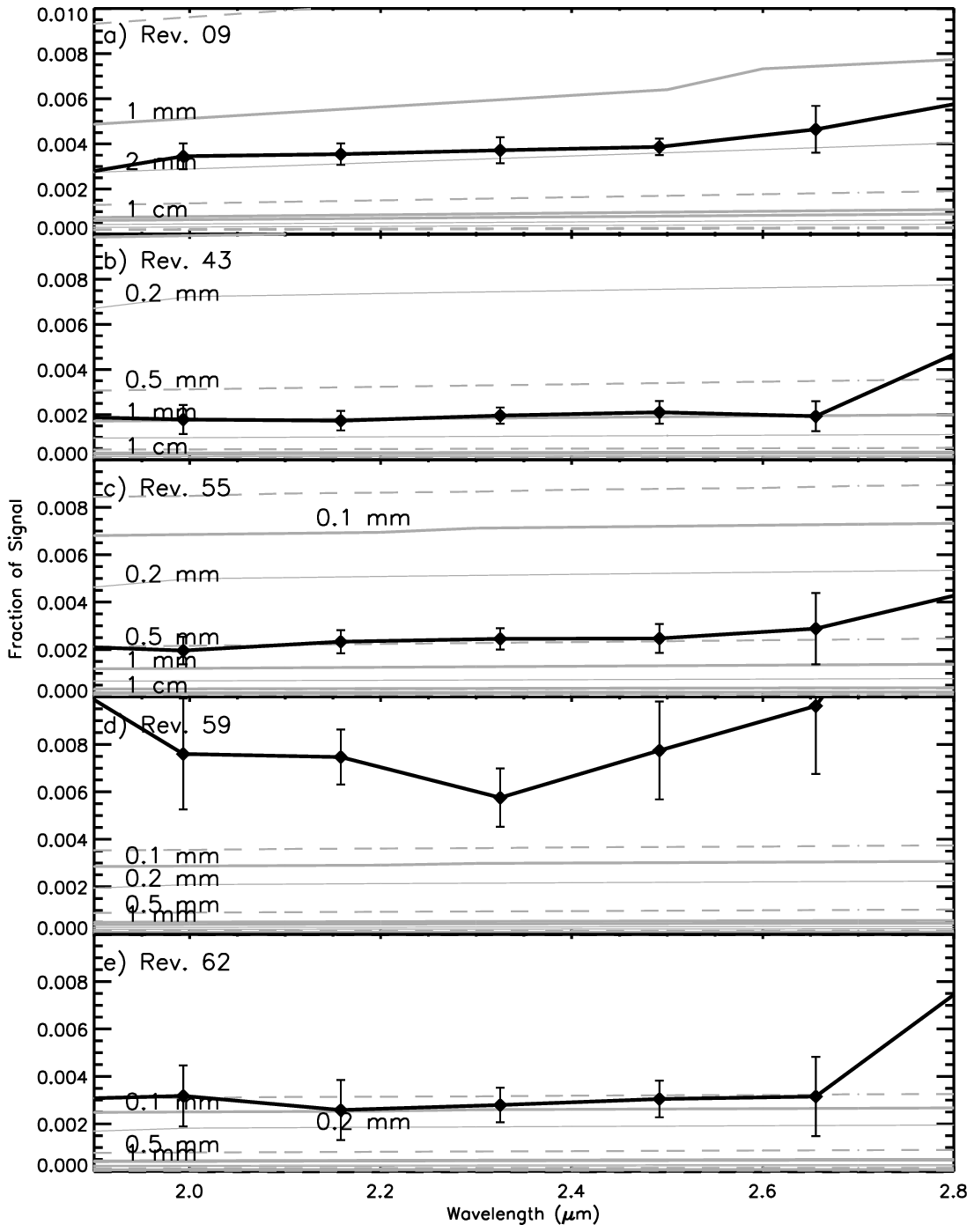} 
  \caption{Five A ring occultations -- Rev. 9 (a), Rev. 43 (b), Rev. 55 (c), Rev. 59 (d), and Rev. 62 (e) -- compared with models ($a_{\max} = 10$m, $q=2.9$, minimum particle sizes from 0.1 mm to 10 cm, and multiple scattering included, but self-gravity wakes not included) with the minimum particle size listed. }
 \label{fig:fractionaringnowakes}
\end{figure}

As before, the values of $f$ were averaged over the entire A ring, and binned spectrally.  Figure \ref{fig:fractionaringwwakes} shows the binned and rescaled measurements of $f$ for five occultations, with representative models.   For a comparison, a wakeless model using the full observed optical depth (but including multiple scattering), is also shown in Figure \ref{fig:fractionaringnowakes}. 

%Enlarge Plot

%Enlarge Plot

Of the five clear positive detections mentioned in Section \ref{sec:specscatter} the diffracted light measurements were larger than we'd expect from models for the Rev. 59 and Rev. 62 occultations.  Below $a_{\min}\approx 100$ microns, the fraction of light removed from the direct signal becomes nearly constant, as the models are no longer dominated by the large-angle `tails' of diffraction from the millimeter-sized and larger particles in the ring.  Rev. 62's measurements only allow an upper limit on $a_{\min}$ to be set, rather than having a value that best agrees with the data, and the data from Rev. 59 are inconsistent with the model entirely for the value of $q$ used.  Omitting the effects of self-gravity wakes, as in Figure \ref{fig:fractionaringnowakes}, changes the minimum particle size corresponding to a given value of $f$, but still cannot reproduce the Rev. 59 observations. 

%Enlarge Plot
 \begin{figure} [htbp]
 \centering 
 \includegraphics[ width=1.0\textwidth]{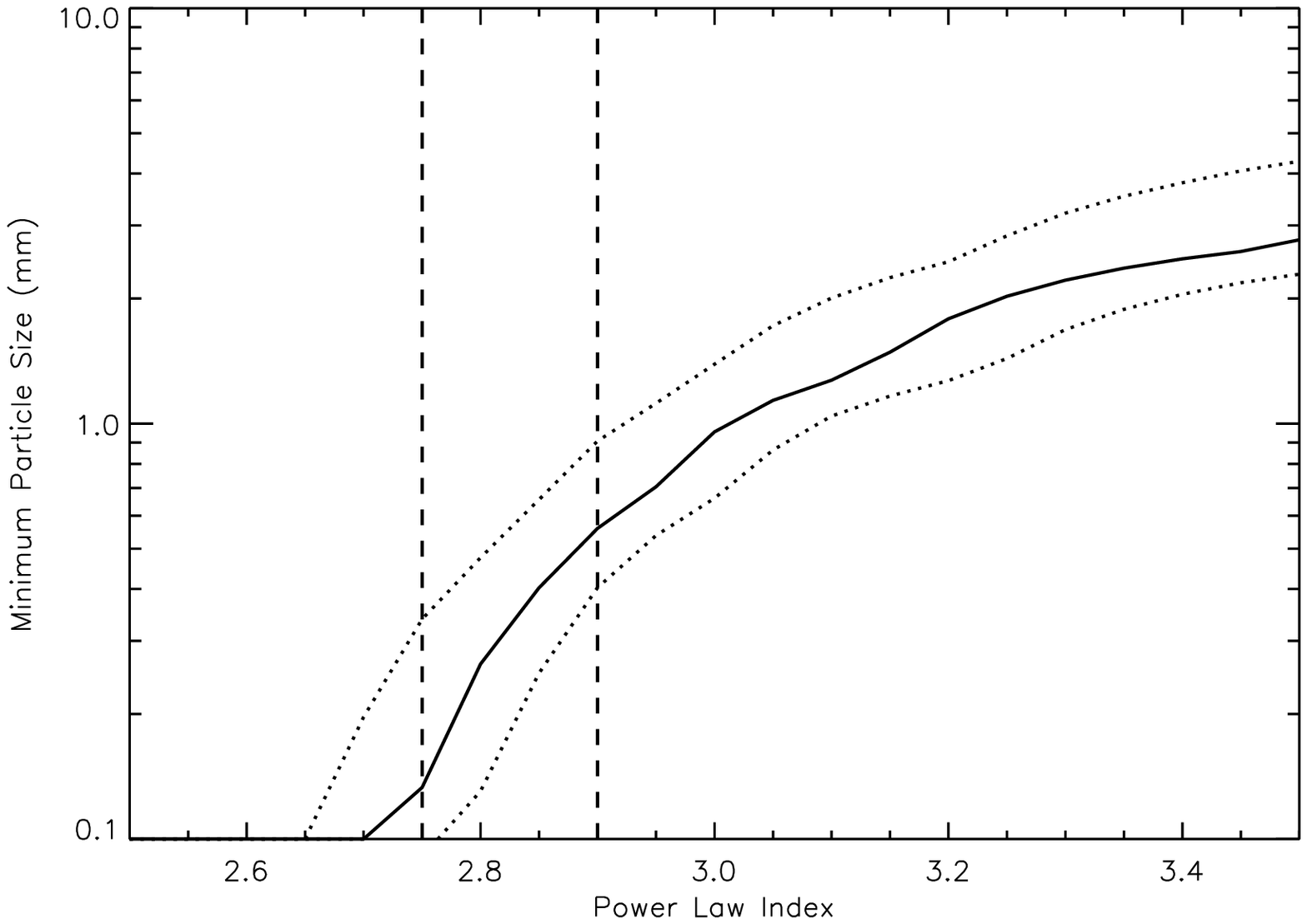} 
   \caption{A plot of $a_{\min}$ as a function of $q$ in the A ring for a wavelength of $2.3\mu$m, assuming wake properties as listed in the body of the text and a maximum particle size $a_{\max} = 10 m$. The function was calculated by taking the scattering fraction from the Rev 09, 43, and 55 occultations, and calculating the $a_{\min}$ for a given $q$ needed to produce the observed scattered light.  A mean was then taken of the three functions. The dotted lines represent 1$\sigma$ errors on the estimates, combining both the differences between the calculated $a_{\min}$s from each occultation, and the errors of each occultation's $a_{\min}$ (calculated from the errors in $f$ calculated from binning nearby wavelengths).  The dashed lines at $q=2.75$ and $q=2.9$ represent previous estimates of the power law index for the A Ring. \citep{1985Zebker1, 2000French1, 2008Marouf1} }
 \label{fig:amodelprofile} 
\end{figure}

To better quantify our results, we again calculated the mean $a_{\min}$ over the three occultations (Revs. 9, 43 and 55) for which a clear detection (rather than an upper bound) was observed, as a function of $q$ from the fraction of scattered light observed at 2.3 microns, just as we did for the C ring.  The results are shown in Figure \ref{fig:amodelprofile}.  Using the diffraction model that accounts for both the effects of self-gravity wakes on optical depth and double and triple particle scattering, we infer that the minimum particle size is $0.56^{+0.35}_{-0.16}$ mm at a power law index of 2.9, the index inferred by the Voyager Radio Science experiment \citep{1985Zebker1}. The shallower $q = 2.75$ power law index observed by \citet{2000French1} lowers the minimum particle size to an upper limit of  $< 0.18$ mm.  Including the Rev 59 and 62 occultations in the mean $a_{\min}$ lowers these values further to $0.38^{+0.27}_{-0.12}$ mm at $q=2.9$, but cannot replicate all the observations using $q=2.75$.

Both the homogenous ring and wake model give a minimum particle size somewhat smaller for expected values of $q$ (between 2.7 and 3.0) than those seen by the Cassini RSS measurements and \citeauthor{2000French1}'s observation of few sub-centimeter-sized particles in the 28 Sgr occultation \citep{2008Marouf1, 2000French1}.  \citet{1985Zebker1} note that the difference in optical depth between that measured at $\lambda$=3.6 cm by Voyager and that measured at 0.5 $\mu$m is large enough to suggest the existence of a substantial population of sub-centimeter sized particles, but a significant difference in optical depth between the 3.6 and 0.9 cm bands in the A Ring was not seen by Cassini RSS occultations\citep{2008Marouf1}, implying few particles in the centimeter size range.  

A major caveat to all of these studies is that none of them accounted for the effects of self-gravity wakes, though \citet{1985Zebker1} and \citet{2008Marouf1} both included analysis of multiple scattering effects.  \citet{2000French1} even notice what could have been a longitudinal asymmetry in optical depth in the A Ring between the $\delta$ Sco and 28 Scr optical depths, but, without a model, chose to adopt a 'fudge factor' to scale the two occultations as best they could.  A model of the A Ring that include self-gravity wakes would lower expected differential optical depths between all wavelengths smaller than the wake size, as a fraction of the optical depth would be caused by the wakes themselves, rather than the continuum of ring particles.  Therefore, a wakeless model would find larger minimum particle sizes for a given differential optical depth than a model that included self-gravity wakes. It is also worth mentioning that our (and others') observations derive distributions for the material in-between the wakes, which may be different in size distribution from the ring as a whole. 

Using the three-occultation mean, our model requires $< 12.1$ \% of the interwake optical depth to be from particles smaller than 1 cm at $q=2.75$, which increases to $20.1^{+4.2}_{-1.2}$ \% for $q=2.9$.  For typical interwake optical depths used earlier ($\tau_G$ between 0.3 and 0.65 in extinction), this gives extinction optical depths due to such small particles of between 0.03 and 0.16, within \citeauthor{1985Zebker1}'s range.  

\section{Conclusions}
When analyzing the solar occultation data recorded by Cassini-VIMS, we observed a small excess of forward-scattered light, once instrumental effects were taken into account.  We believe this to be due to diffraction by small particles in the rings and have used it to estimate minimum particle sizes, assuming a power law index, $q$, and maximum particle size from previous work \citep{1985Zebker1, 2000French1, 2008Marouf1}.

Among the three C Ring solar occultations in which a clear positive excess was measured, a minimum particle size of $4.1^{+3.8}_{-1.3}$ mm is inferred for a canonical value of $q = 3.1$.  For a wider range of likely $q$s, the data still indicate a minimum particle size between 3 and 10 mm.  This is somewhat larger than the $a_{\min}\approx 4$ mm measured by \citet{2008Marouf1} using the Cassini Radio Science experiment, and it's possible this could be due to a radial variation of minimum particle size in the C Ring, as the chord occultations (Rev. 62 and 65) show a larger minimum than the Rev. 09 radial occultation.  Further work would be required to confirm  such a variation. 

In the A Ring observations, multiple-particle scattering produces a non-negligible effect due to the larger optical depths involved, and must be taken into account to explain the larger-than-expected amount of scattered light seen.  The effects of the A Ring's self-gravity wakes on the amount of scattering are more complicated, but are clearly seen in optical depth measurements of the A Ring from both these solar occultations and other data sets (such as stellar occultations).  The shallow power law indices of $q = 2.75$ found by \citet{2000French1} and \citet{2008Marouf1} require a very small $a_{\min}$ of $< 0.34$ mm to explain our observations, even accounting for multiple scattering and self-gravity wakes.  Raising the power law index to $q=2.9$ as measured by the Voyager radio occultations \citep{1985Zebker1} still requires particles of  $0.56^{+0.35}_{-0.16}$ mm to explain the amount of scattered light measured by our solar occultation observations.  These numbers appear to be inconsistent with estimates of a lack of material smaller than one centimeter advanced by \citet{2000French1}, but the shallow power law and amount of material sequestered in self-gravity wakes may mean the optical depth required in particles smaller than 10 mm could be as small as $\tau = 0.03$ in extinction. This may render our data consistent with this lack of optical depth variation with wavelength seen in radio occultations, especially when the effects of self-gravity wakes are taken into account.

We were also able to constrain the fraction of free-floating ice grains smaller than $100 \mu$m in the A ring to be $\le 5\%$, assuming a dust size distribution similar to the F Ring.  The fraction within the C ring was even smaller; $\le 1.4\%$. Regardless of their minimum particle sizes, it is clear that the A and C Rings lack the persistent icy dust that is a strong feature of the F Ring.   

\appendix
\section{Phase Functions}
\label{append:theory}

For a single-size particle distribution, the forward-scattering, or diffraction, phase function is given by \citep{1980Liou1}
\begin {equation}
\label{eq:singleP_append}
P\left(\theta\right) = \left[\frac{2 J_1\left(z\right)}{\sin \theta}\right]^2
\end{equation}

where we introduce the dimensionless variable $z = 2 \pi a \sin \theta / \lambda$, $a$ being the radius of the particles and $\lambda$ being the  wavelength observed. $J_1\left(z\right)$ is the first-order Bessel function of the first kind.  Integrating Equation \ref{eq:singleP_append} over a truncated power law distribution of particle sizes, $dn/da = n_0\left(a/a_0\right)^{-q}$, where $a_{\min}\le a \le a_{\max}$ and $n_0$ and $a_0$ are constants that can be folded into the value of $\tau$, we find 

\begin {equation}
\label{eq:powerP_append}
\overline{P}\left(\theta\right) = \frac{4}{\alpha} \sin^{q-5}\theta \int_{z_{\min}}^{z_{\max}}z^{2-q} J_1\left(z\right)^2\,dz
\end{equation}

where 

\begin {equation}
\label{eq:alphadef}
\alpha=\left\{ \begin{array}{ll}
\ln \frac{a_{\max}}{a_{\min}} & q = 3\\
\frac{x_{\max}^{3-q} - x_{\min}^{3-q}}{3-q} & q \ne 3 \\
\end{array} \right.
\end{equation}

The usual dimensionless size parameter $x$ is defined by $x = 2 \pi a / \lambda$, with subscripts denoting the limiting values of $a$.

The mean phase function (Equation \ref{eq:powerP_append}) can be conveniently approximated in different limiting cases, as the full function can be computationally expensive to integrate.  The limiting cases are set by the relevant angles in the problem, which are determined by the ratio of particle size to wavelength (as quantified by $x$).  Let the minimum characteristic diffraction angle -- the angle where the largest particles will be diffracting light -- be $\theta_1 = \pi x_{\max}^{-1}$.  Similarly, we define the maximum characteristic diffraction angle (where the smallest particles will be diffracting light) as  $\theta_2 = \pi x_{\min}^{-1}$.  

Two angles give us three cases to consider, but only two are of real interest in this case.  Small-angle diffraction -- where the angles we observe at are all smaller than $\theta_1$ -- isn't relevant here, as the upper boundary of the ring particle size-distribution in the A and C Rings extends to 5m in radius §\citep{1985Zebker1}, and at near infrared wavelengths (0.9 to 5.2 $\mu$m), this corresponds to a $\theta_1$ of tenths of microradians.  Thus we either have a case of medium-angle diffraction (the angles we observe are between $\theta_1$ and $\theta_2$) or large-angle diffraction (all angles observed are larger than $\theta_2$).  

The value of $\theta_2$ is unknown, because the minimum particle size is the quantity we are trying to measure. Given that the size of one VIMS pixel -- and coincidentally the solar radius at 9 AU -- is 0.5 milliradians on the sky, our data will be most sensitive to diffraction by particles with $x \lesssim 6000$, or, at 2 microns wavelength, particle sizes of 2 millimeters or less.  Barring a much-lower-than-expected minimum size cutoff, the large-angle scattering case will be most relevant, though we will include the medium-angle case in our calculations to account for the possibility of free-floating particles from $\sim$100 $\mu$m to $\sim$2 millimeters.  

For the large-angle case, (i.e. $\theta\gg\theta_2$), all particles are scattering most of their light at angles smaller than those we are measuring.  Thus the bounds on the integral of Equation \ref{eq:powerP_append} are both much larger than unity.  We can then use the approximation $J_1(z) \approx \sqrt{2/\pi z} \cos ( z - 3\pi/4)$, giving

\begin{equation}
\label{eq:largebesselapprox_mid_append}
\overline{P}\left(\theta\right) \approx \frac{4}{\pi\alpha}\left(\sin\theta\right)^{-3}\frac{x_{\min}^{2-q}-x_{\max}^{2-q}}{q-2}.
\end{equation}

Because the particle size distribution is very broad (remember we're dealing with particles with radii from millimeters to meters in size), we also know that $x_{\max} \gg x_{\min}$, and both are very large.  So, a further approximation is to drop the $x_{\max}^{2-q}$ term (which will be very small as long as $q > 2$), which leaves the simpler expression

\begin{equation}
\label{eq:largebesselapprox_final_append}
\overline{P}\left(\theta\right) \approx \frac{4}{\pi\alpha}\left(\sin\theta\right)^{-3}\frac{x_{\min}^{2-q}}{q-2}.
\end{equation}

In the case of medium angle diffraction (i.e. $\theta_1\ll\theta\ll\theta_2$), we again use a broad particle size distribution to approximate a phase function.  Because of this distribution and an angle ($\theta$) that is between the minimum and maximum characteristic diffraction angle, we are mostly sampling light neither from the smallest nor the largest particles, but from medium-sized ring particles that have that characteristic diffraction angles.  Because $\theta$ is much smaller than the maximum ($\theta_2$), we can assume that $z_{\min} =\pi \sin \theta/\theta_2$ is much less than unity, and because $\theta$ is much larger than the minimum ($\theta_1$), we can assume that $z_{\max} =\pi \sin \theta/\theta_1$ is much greater than unity.  We can then approximate the integral in Equation \ref{eq:powerP_append}, as covering the full range of positive values of $z$, from zero to infinity, as most of the power is around $z \approx 1$.  This leads to a constant that is only dependent on $q$, allowing the integral to be calculated once per $q$.  Thus, we have the approximation

\begin{equation}
\label{eq:intbesselapprox_append}
\overline{P}\left(\theta\right)\approx \frac{4}{\alpha}\left(\sin\theta\right)^{q-5}\mathcal{J}_0^\infty\left(q\right),\,\theta_2\le\theta\le\theta_1.
\end{equation}

The $\mathcal{J}_0^{\infty}(q)$ in Equation \ref{eq:intbesselapprox_append} is shorthand for $\int_0^{\infty}z^{2-q}{ J_1\left(z\right)}^2\,dz$.  It is nearly constant over the range of $2\le q \le 5$, except when $q$ approaches 2 or 5.  Previous studies indicate that $q$ is between 2.7 and 3.1 within the main rings, giving $J_0^{\infty}\approx0.5$ (\citealt{1985Zebker1}, \citealt{2000French1}, \citealt{2009Cuzzi1}).

\bibliographystyle{model2-names}
\bibliography{sources}
\end{document}